\documentclass[usenatbib,twocolumn]{mn2e}

\usepackage{graphicx,amsmath,amssymb,enumitem,verbatim,sidecap}
\begin{document}

\title[]{Multiwavelength cluster mass estimates and machine learning}

\author[]{ J.D.Cohn${}^{1,2}$\thanks{E-mail: jcohn@berkeley.edu} and Nicholas Battaglia${}^{3}$\\
${}^1$Space Sciences Laboratory 
  University of California, Berkeley, CA 94720, USA\\
${}^2$Theoretical Astrophysics Center,
  University of California, Berkeley, CA 94720, USA \\
${}^3$Cornell University, Ithaca, NY 14853, USA \\}
\maketitle
\begin{abstract}
One emerging application of machine learning methods is the inference of galaxy cluster masses.
In this note, machine learning is used to directly combine five simulated multiwavelength measurements in order to find cluster masses. This is in contrast to finding mass estimates for each observable, normally by using a scaling relation, and then combining these scaling law based mass estimates using a likelihood.  We also illustrate how the contributions of each observable to the accuracy of the resulting mass measurement can be compared via model-agnostic Importance Permutation values.  Thirdly, as machine learning relies upon the accuracy of the training set in capturing observables, their correlations, and the observational
selection function, and as the machine learning training set originates from simulations, two tests of whether a simulation's correlations are consistent with observations are suggested and explored as well. 
\end{abstract}
\begin{keywords}

\end{keywords}

\section{Introduction}
Galaxy clusters are the most massive bound structures in the universe and are often complex.  This is in part due to their 
tendencies to be
triaxial, to lie in an anisotropic environment at intersections (nodes) of the cosmic
web, and to be recently formed.  
A cluster's host halo can be characterized by its mass, which characterizes many properties of the gas, galaxies and other quantities within
it \citep[for a recent review, see][]{WecTin18}.  Clustering with other
halos is also most affected by a cluster's mass \citep{Kai84,ColKai88,Efs88,MoWhi96},
although there are secondary (``assembly bias'') effects which are
also under study, both for clustering and for other cluster properties, such as number of galaxies within.  There also is interest in
cluster masses due to the sensitivity of the cluster mass
distribution to cosmological parameters
\citep[e.g.,][]{Voi05,AllEvrMan11,BorKra11}.

A particular cluster boundary definition 
specifies a cluster's halo and its mass corresponding to the particular definition.
In a simulation, a definition of where a cluster halo begins and where
its environment ends (given the physics included in the simulation) can be precisely applied to find the mass, especially using the available three dimensional spatial and
velocity information.  Observationally,
it is much harder to disentangle a cluster from its
surroundings and from direction dependent effects, and to account for confounding dynamical effects.  For example, cluster mass can be indicated by the number of galaxies the cluster hosts
(richness), but galaxies are observed in redshift space and thus can
lie several Mpc from the cluster but appear to be within it.  The hot
cluster gas scatters CMB photons en route to us via the
Sunyaev-Zel'dovich effect \citep{SunZel72}, effectively leaving a
shadow on the cosmic microwave background or CMB, but other mass along
the line of sight to the cluster can also scatter these photons and
thus modify the signal.  Weak lensing shear captures all the mass in
the cluster, but only as part of capturing all the mass between the galaxy
being sheared (located behind the cluster) and us (in front of the
cluster).  Although X-ray measurements tend to only track the hot gas
within the cluster, the dynamical state of the gas can significantly
modify estimates, and even for X-ray, projection effects can contaminate the sample \citep[e.g.,][]{Rametal19}.  There is also
  a suggested dynamic cluster 'edge' definition based upon velocities and positions of
  galaxies in a cluster, via their splashback
  \citep[for example,][]{AdhDalCha14,MorDieKra15} or backsplash
  \citep[for example,][]{GilKneGib05} radius; see, for example, \citet{Shi19} for a recent measurement.  These dynamics are used in the caustics \citep{Rin03}
  method to find cluster masses, which has scatters due to interlopers and line of sight variations in velocity dispersions \citep[e.g.][]{GifMilKer13}.

Multiwavelength measurements of clusters can be combined to get a more complete picture of any given individual cluster or of a distribution of clusters.  Often, the route for determining a cluster's mass is to find a mass-observable relation
for each cluster property, separately,
e.g. galaxy richness or SZ flux.  These fits (usually power law scaling relations) convert observables into mass estimates.  For
multiwavelength measurements, these power law
based mass estimates plus their estimated errors are combined in a likelihood to
estimate the underlying cluster mass \citep[see][for a detailed review and discussion]{AllEvrMan11}.  However, as the scatter between the true and
estimated mass for each observable is often due to the cluster's
triaxiality, anisotropic environment and other physical cluster
properties, the mass estimate errors using different observables can be correlated.
These mass error correlations should be represented faithfully in the analysis, or they can
impose a bias 
\citep[see, for example,][for more details
and examples of
correlations]{AllEvrMan11,Ryk08,Sta10,WCS,Ang12,ShiNagLau16}. 
Correlations are also
possible between the scatter in the observable used to
select the clusters and observables used to measure the cluster masses.  To summarize: observables are each converted to cluster mass via an observable specific scaling relation (usually a power law), and then combined using correlations whose form is either marginalized over or estimated from simulations.

In this note, machine learning is suggested as a method to obtain multiwavelength
mass estimates of a cluster directly from the observational quantities, skipping the intermediate step of matching observations to average relations for mass (of a predetermined form, i.e.,
scaling relations, plus estimating their correlated mass errors) and instead going straight to the mass prediction of the
combined observations.  One reason one might expect this approach to be
successful is that scatter around the true mass for multiwavelength observations can be analyzed
using principal component analysis \citep{NohCoh12}, where it can be
seen that the leading contribution to the scatter for each observable tends to be occur in a certain
combination with that of other observables.

This suggested approach follows the philosophy of
\citet{Ho19} and earlier papers \citep{Nta15,Nta16,Nta17}.  There, galaxy cluster masses are
estimated by using cluster galaxy velocities, but instead of turning
all the cluster velocities into a single velocity dispersion and mapping the dispersion 
into a mass using a mean relation (approximate fit), the individual velocities are given to a
machine learning algorithm, to encode the more detailed relations between the
different galaxy velocities, their angular positions and the cluster final mass.  Additionally, \citet{ArmKayBar19} combine, with no measurement errors (but fixed length cylinders to include some projection effects), several X-ray, SZ, galaxy angular position, velocity dispersion and count properties to obtain machine learning cluster masses.  They consider a large range of features and include the gradient boost method as we do, below, but not the other two random forest related methods, which we found had better success.  Amongst the methods they tested, they found the most success with ordinary linear regression and variants of ridge regression.

Here, using a simulation which has several different multiwavelength mock
cluster observations, machine learning is used
to 
\begin{enumerate}
\item calculate the mass directly from the combination of the
different multiwavelength
observables, without going through a mass estimate for each,  
\item compare subsets of multiwavelength measurements to each other to identify which combinations give the smallest scatters and biases in predicting mass (within the measurement approximations which are made in the simulation), and 
\item
test for whether a given simulation correctly captures
correlations between different observables (again, within the context of a specific mass-observable relation given by the simulation).  
\end{enumerate}
Having all multiwavelength mock observations made on the same
simulation guarantees a shared model for all the observables which
are being compared.  However, this suggested approach is also limited in its
usefulness by the accuracy of the simulation (and how well this
accuracy is understood); hence the interest in identifying tests for
correlations such as are discussed below.  This accuracy is also important for the usual scaling relations method, if a simulation is used for calibrating the scaling relation.

Again, using the various observations together in a machine learning algorithm is to be contrasted to mapping each individual observable's measurement
to mass via a predetermined scaling relation plus scatter (from a simulation) and then combining probabilities.
The hope is that the simultaneous fit might include more information
from the relations between the multiwavelength measurements and thus improve the accuracy.
In section \S\ref{sec:methods}, the simulation measurements and
machine learning methods are
described.  In section \S\ref{sec:masstests}, mass
estimates are made, and errors characterized.  In \S\ref{sec:validate}, a few ways of testing a
simulation relative to observations are suggested and explored with a simulation (relative to a simulation with shuffled observables, i.e., with artificially suppressed correlations). \S\ref{sec:discuss} concludes.

\section{Methods and data}
\label{sec:methods}
\begin{figure*}
\begin{center}
\resizebox{6.in}{!}{\includegraphics{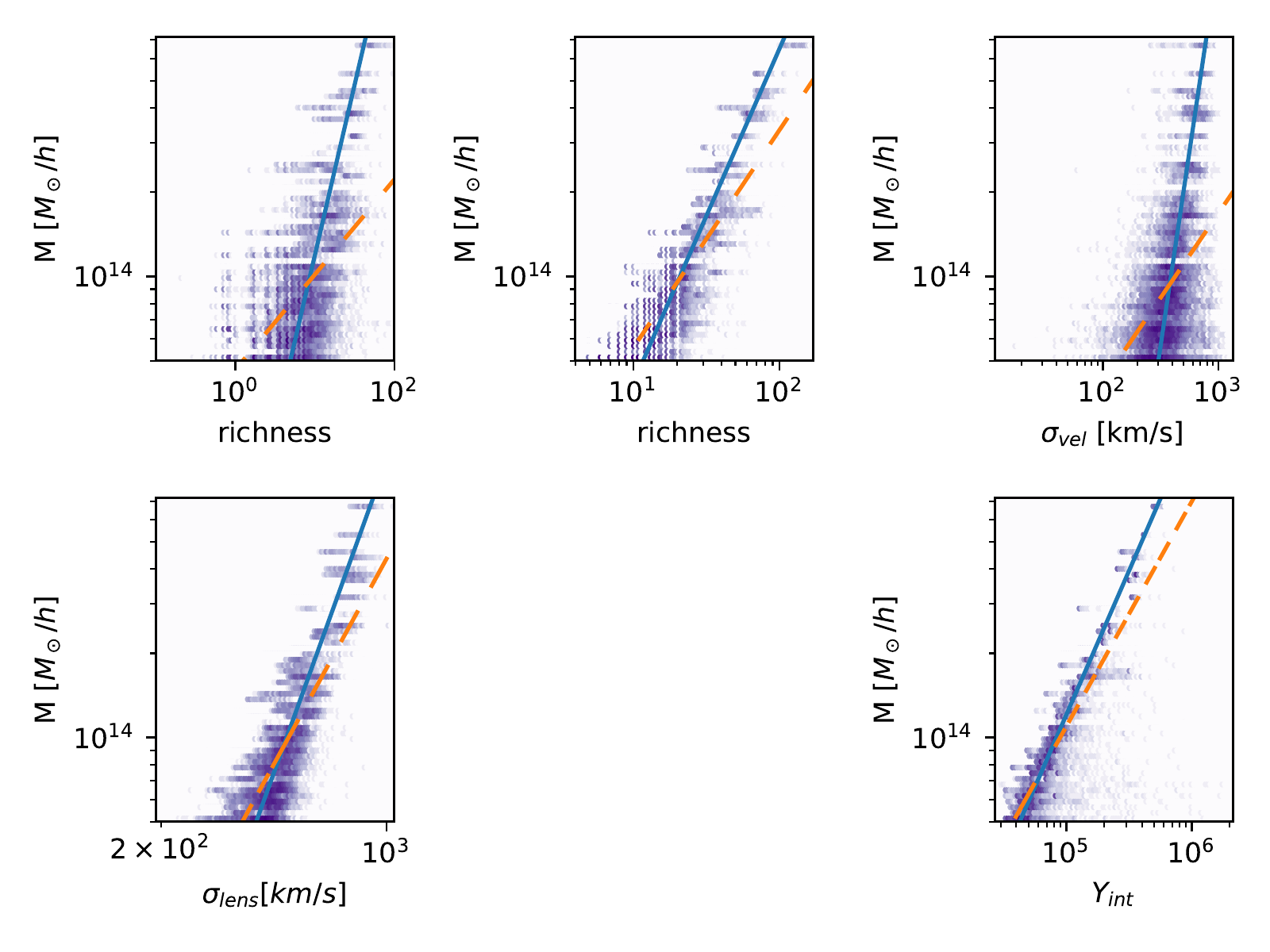}}
\end{center}
\caption{Distribution of each observable for all the lines of sight of all halos with $M> 5 \times 10^{13} h^{-1} M_\odot$ in the box.  Solid lines in each log-log plots show scaling relations found by binning each observable in 20 bins of log mass, and then doing a least squares fit, while dashed lines correspond to a direct least squares fit.}
\label{fig:observables}
\end{figure*}

We use a dark matter only simulation 
in a periodic box of side $200 $ Mpc $h^{-1}$ with $1500^3$ particles, 
evolved using the TREEPM \citep{TreePM} code, and provided by Martin
White.  Cluster observables are assigned as in
\citet{WCS}, hereafter WCS,  which can be consulted for details beyond those found below.  Although the sample used for analysis in WCS itself has a slightly larger box with higher resolution, its cluster measurements did not extend to low enough mass to reasonably expect completeness for some cuts considered here.  Combining the two samples did not seem appropriate given their different resolutions and cosmologies. 

The simulation background cosmological parameters are $h=0.72$, $n=0.97$,
$\Omega_m=0.25$, and $\sigma_8=0.8$, 
slightly off from the best
estimated current cosmology \citep{Planck16}, but close enough for the concerns
of interest in this note.
There are outputs at 45 times equally spaced in $\ln(a)$ from $z=10$
to $0$.  Halos are found using a
Friends of Friends (FoF) halo finder \citep{DEFW}, with linking length $b=0.168$ times
the mean interparticle spacing.  Masses quoted below are FoF masses.
 As the box is fairly small, only $z=0.1$ is considered, with
117 halos with $M> 10^{14} h^{-1} M_\odot$ and 339 halos with $M>5
\times 10^{13} h^{-1} M_\odot$.  Mock observations are made of five observables.  For each observation, the cluster is placed at the center of the box.  To get more statistics, and to explore the effect of anisotropy, 
triaxiality, and their correlations,
each cluster is "observed" via all five methods below, from 96 different
isotropically distributed
directions on a sphere.  The methods are:
\begin{itemize}
\item $N_{\rm red}$: red galaxy richness based on the older MaxBCG mass estimator
  \cite{maxBCG}; colors are assigned using \citet{SkiShe09}, with redshift evolution as detailed in appendix B2 of WCS.
\item MF: richness from matched filter based upon \citet{Yan07}.  This gives each galaxy at position $(R,\Delta r_\parallel)$ from the cluster center 
a weight made up of a projected NFW halo profile \citep{NFW} in the plane of the sky, times a Gaussian in $\Delta r_\parallel$ along the line of sight, with the galaxy kept if this weight exceeds a threshold value of 10.  The details are in equations 7-11 of \citet{Yan07}, with redshift $\Delta z$ replaced with real space $\Delta r_\parallel$, and $H_0$ replaced with $H$.  The matched filter dispersion and characteristic radius are based on the true halo mass.
\item vel: velocity dispersions, calculated using the method descibed in detail in WCS in \S 5.1.
\item $Y_{\rm int}$: Sunyaev-Zel'dovich decrement $y$ \citep{SunZel72}. Each particle is given the mean temperature associated with the velocity dispersion of its host halo.  For each halo center, the integration of the SZ signal is done within a disk of radius $r_{180b}$ perpendicular to the line of sight, through the entire box along the line of sight (apodized as in WCS Eq.~1 to avoid a sharp boundary cutoff).  
\item WL shear: fit to a dispersion $\sigma_{\zeta}$ using the $\zeta$ statistic and a singular isothermal sphere profile.  These are described in \S 3.3 of WCS.  Roughly, $\zeta$ (WCS Eq.~2) corresponds to subtracting the projected surface density in an annulus from that of a disk to decrease the contributions from material uncorrelated with the cluster.  Then $\zeta$ is converted to $\sigma_\zeta^2$ by multiplying factors drawn from assuming some particular cluster profile, Eq.~3 in WCS. An isothermal sphere was used; similar results were obtained as profiles and annuli parameters were varied.
\end{itemize}
See WCS and the other papers referenced above for further detail on these techniques.

The distribution of the resulting observables as a function of mass is shown in Fig.~\ref{fig:observables}.  Lines are fit either using a least-squares fit (dashed line) or binning in 20 log $M_{\rm true}$ bins and finding the median observable in each and fitting to those, dropping bins with no clusters in them.
These are meant to be mock observations, that is, relying primarily only on information
available to an observer (rather than the three dimensional
velocities and positions, exact gas content, etc., which a simulator might be able to access). However, some of the mock observations have advantages over real sky observations:  using true $r_{180b}$ and not including any survey dependent residuals from foreground cleaning (SZ), using real space along the redshift direction and true halo mass matched filter dependence (MF), and neglecting survey dependent photometric redshift magnitude and shape errors and galaxy shape errors (WL shear). 
Also, the MaxBCG richness is not in
standard use anymore, however, its replacement in several new and upcoming large surveys \citep[{\tt redmapper},][]{redmapper}
does not yet have a halo model or other description which would allow
its richness prediction to be reliably calculated for this sample.  (There are many other richness
estimators possible, see \citet{Old15} for a comparison of several of
them, applied to two blinded
data sets.)

For machine learning, for each mock multiwavelength observation, the
input variables are the five observables ($N_{\rm red}$, MF, vel,
$Y_{\rm int}$,WL shear) for each cluster, along a single line of sight.
The output is the true mass $\log_{10} M_{\rm true}$ for the training set, and the
 mass prediction $\log_{10} M_{\rm pred}$ for the testing set.  The training set is a random
selection of the set of cluster observations (1/10 of the sample), with testing done on the
other 9/10.  The machine learning programs used
are {\tt GradientBoostingRegressor}, 
{\tt ExtraTreesRegressor} and {\tt
  RandomForestRegressor} from sklearn in python.
  
  \noindent The exact
  calls were: 
  \begin{itemize}
  \item {\tt
    GradientBoostingRegressor(loss='ls', learning\_rate=0.1,
    n\_estimators=100, subsample=1.0, criterion='friedman\_mse',
    min\_samples\_split=2, min\_samples\_leaf=1,
    min\_weight\_fraction\_leaf=0.0, max\_depth=3,
    min\_impurity\_split=None, init=None, random\_state=None,
    max\_features=None, alpha=0.9, verbose=0, max\_leaf\_nodes=None,
    warm\_start=False, presort='auto')} 
 \item {\tt
  ExtraTreesRegressor(n\_estimators=700, min\_samples\_split=5,
  n\_jobs=-1)}
  \item {\tt
  RandomForestRegressor(n\_estimators=300, n\_jobs=-1,
 min\_samples\_split=5)}
 \end{itemize}
 The {\tt GradientBoostingRegressor} uses default values for the arguments, while {\tt ExtraTreesRegressor} and {\tt RandomForestRegressor} follow 
 \citet{KamTurBru16a,KamTurBru16b} for parameters in the
 function calls.  
 
 There are several detailed descriptions of these methods in the literature, in particular \citet{Bre01} and \citet{GeuErnWeh06} for Random Forests and Extra Trees (see also the descriptions for astronomers in \citet{KamTurBru16a}, \S 2.4.2 and 2.4.3).  The books by \citet{HasTibFri09} and \citet{Ger17} have pedagogical approaches, with the latter oriented around scikit.learn in particular.
 
Heuristically, Random Forests and Extra Trees create regression trees based upon a training sample, splitting the sample according to one feature at a time until some condition is met.  
Both methods choose the feature for each subsequent split by taking a random subset of the features (in the case here, the features are the 5 observables,  $N_{\rm red}$, MF, vel, $Y_{\rm int}$ and WL shear) and testing possible bifurcations, one for each of the features in the random feature subset.  The “best” bifurcation is the one which roughly has the least final variance, where variance is calculated with respect to the mean in each branch, over all the elements in each branch, with a weighting determined by the fraction of elements in that branch. After applying this "best" bifurcation, if the stopping condition is not met, the process repeats to create the next bifurcation of the tree.  Both methods create an ensemble of trees and combine the results to reduce overfitting and improve accuracy.

The big differences between the Random Forests and Extra Trees are that the training set is used differently and the bifurcations at each step in the regression tree are chosen differently.  In Random Forests, the full training set is not used for every tree in the ensemble. Instead, a bootstrap sample of the training set is used, with a different bootstrap sample for each tree that is in the combined final ensemble. Secondly, when the bifurcations are made, the “best” bifurcation is found separately for each feature (with “best” as above), and then the resulting best bifurcations of each feature are intercompared.  The best of the best bifurcations is chosen to make the next tree bifurcation.  For Extra Trees, instead, the whole training set is used to train each tree.  Secondly, for bifurcations, a random bifurcation is chosen for each of the features being compared.  These random bifurcations of each feature are compared to each other (rather than the best bifurcations of each feature in the Random Forest), but again, the “best” one amongst the features is taken.  The Extra Trees method is much faster, as the “best” bifurcation is not first found for each feature separately, before comparing the separate proposed feature bifurcations to each other. It is not always evident which one will do a better job on a given problem until they are both tried.\footnote{{\tt DecisionTreeRegressor} is also used in \citet{KamTurBru16a}). It fits to a single tree rather than an ensemble.  It fails to provide reasonable scatter for the sample here, due to the small number of distinct true halo masses, especially at the high mass end.}

The gradient boost method uses an ensemble of trees, as well, but instead of taking a tree ensemble with different trees having different random training samples or bifurcations, it creates a sequence of trees, where each tree fits the negative gradient of the loss function of the previous tree.  The loss function chosen here is least squares. 
  
These particular methods were chosen due to their familiarity for the authors; an exhaustive comparison of all the methods available in scikit.learn \citep{scikit} or elsewhere, or possibilities with custom made ML codes, was not undertaken.
 A more extensive comparison and exploration would be interesting to do for a machine learning study, given the initial success with these three "out-of-the-box" methods. 

\section{Results}

\subsection{Predicting mass}
\label{sec:masstests}
\begin{figure}
\begin{center}
\resizebox{3.3in}{!}{\includegraphics{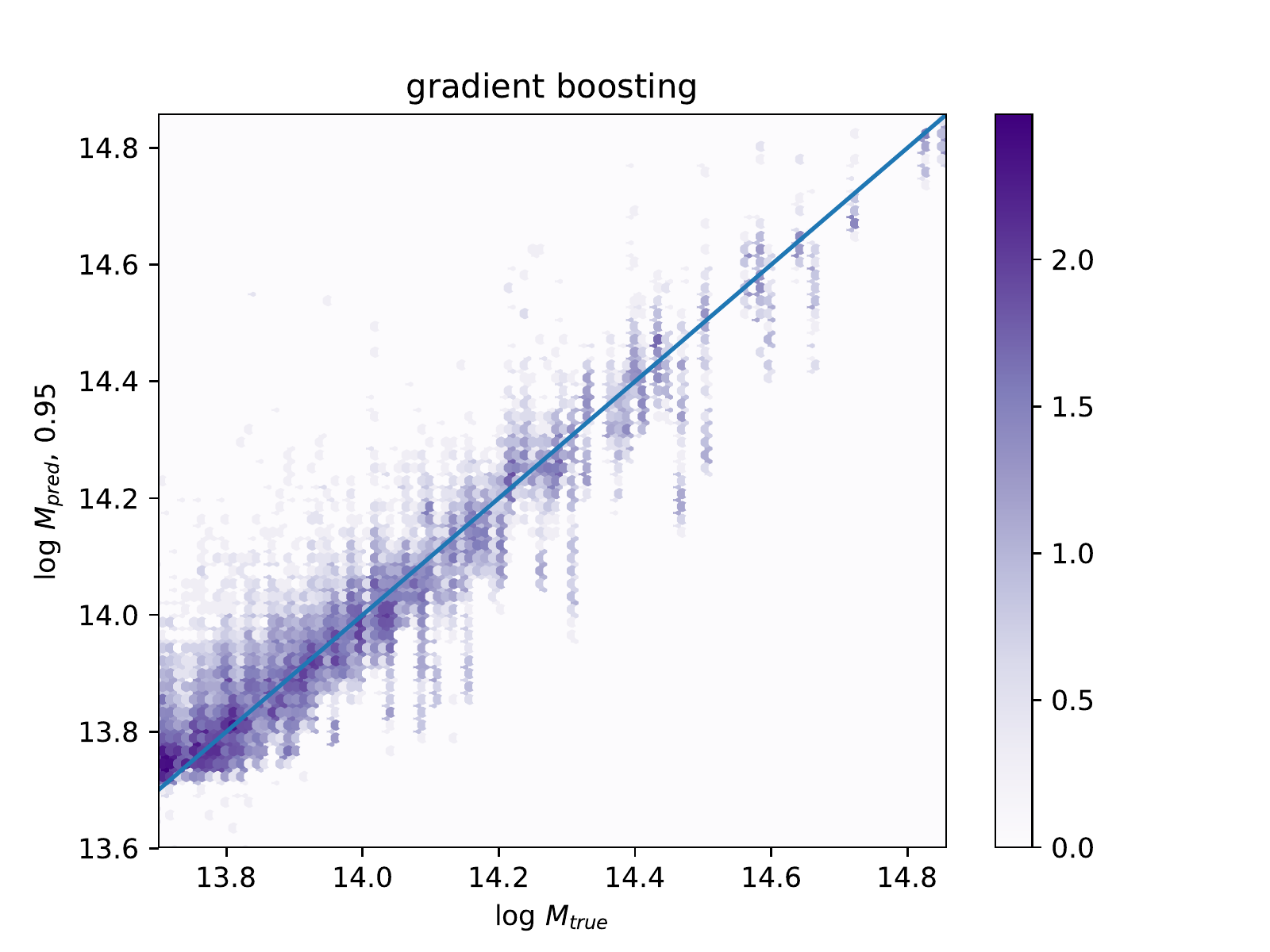}}
\resizebox{3.3in}{!}{\includegraphics{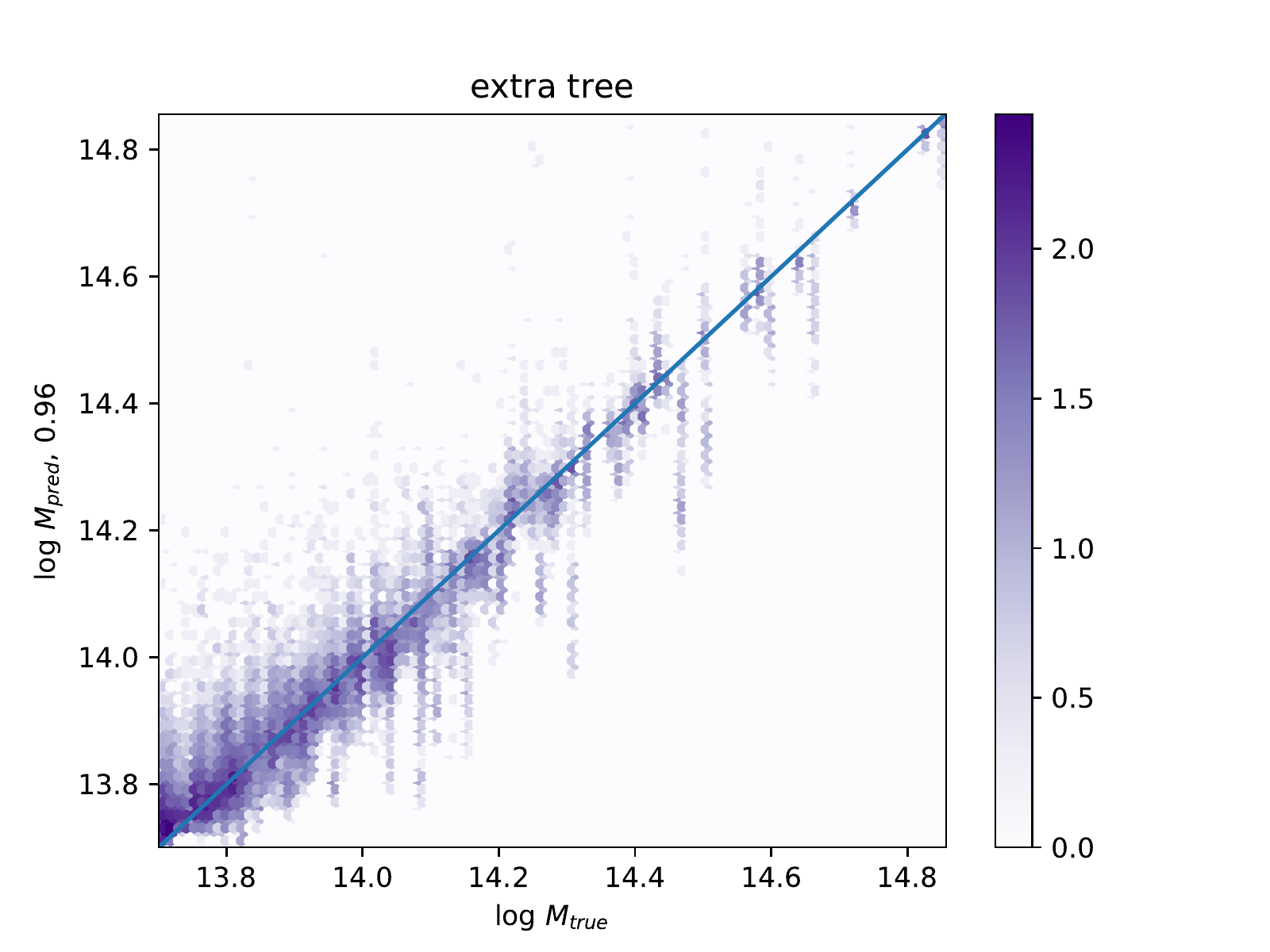}}
\resizebox{3.3in}{!}{\includegraphics{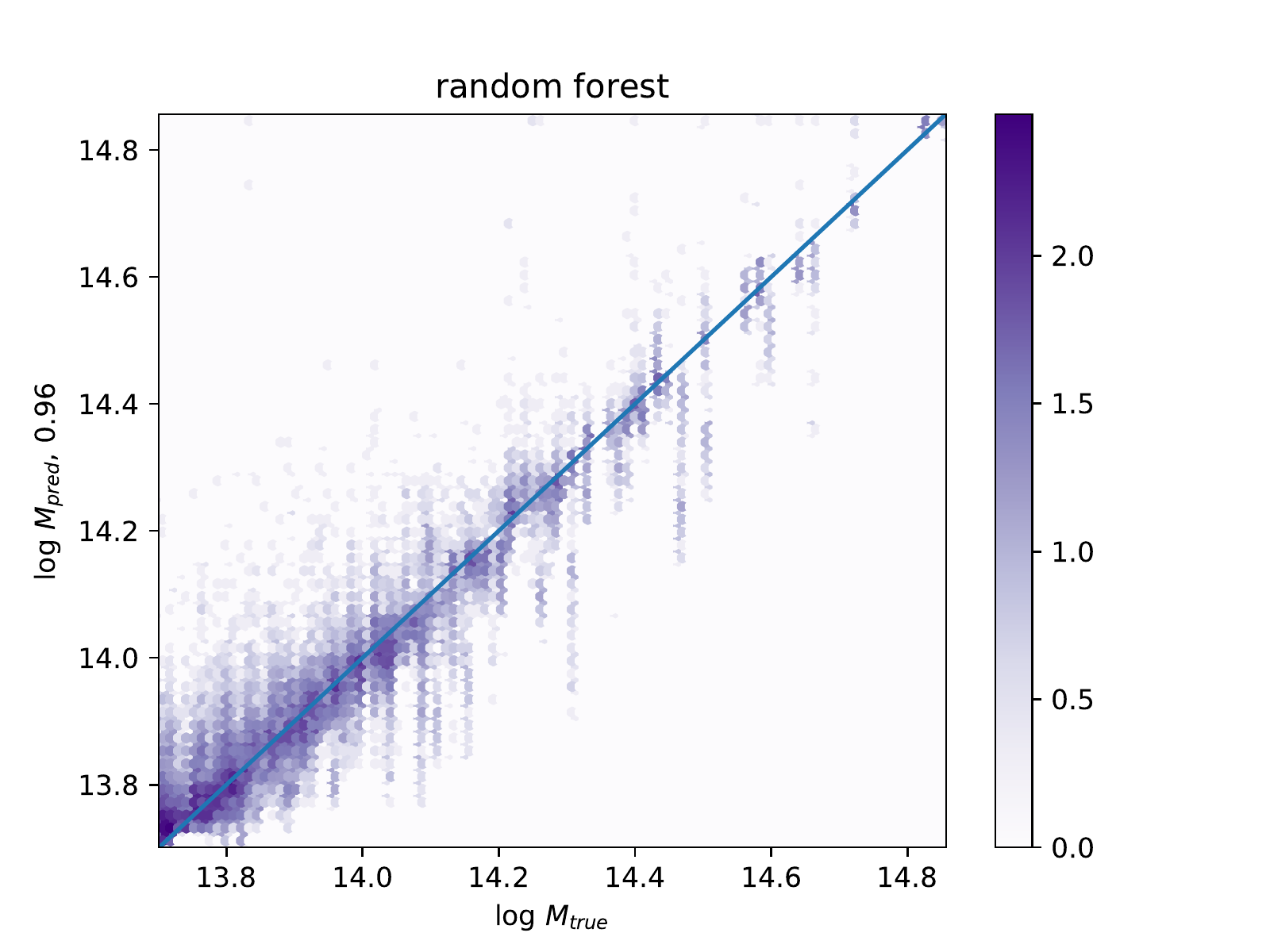}}
\end{center}
\caption{True and predicted $\log_{10} M/(h^{-1}M_\odot)$ for several
  different machine learning algorithms, Gradient Boost, Extra Trees and
  Random Forests, top to bottom.  The correlation between true and
  predicted values is shown on the $y$-axis.  Stripes at fixed
  $M_{\rm true}$ appear because each cluster appears 96 times,
  observed from 96 different angles.  The color scale corresponds to $\log_{10}
  N_{\rm clusters}$.}
\label{fig:corrlns}
\end{figure}

The correlation between $\log_{10} M_{\rm true}$  and $\log_{10} M_{\rm pred}$  for the three
 machine learning methods are
shown in Fig.~\ref{fig:corrlns}, and lie in the range 95\%-96\%,
listed on the $y$-axis for each panel.
\begin{figure}
\begin{center}
\resizebox{3.3in}{!}{\includegraphics{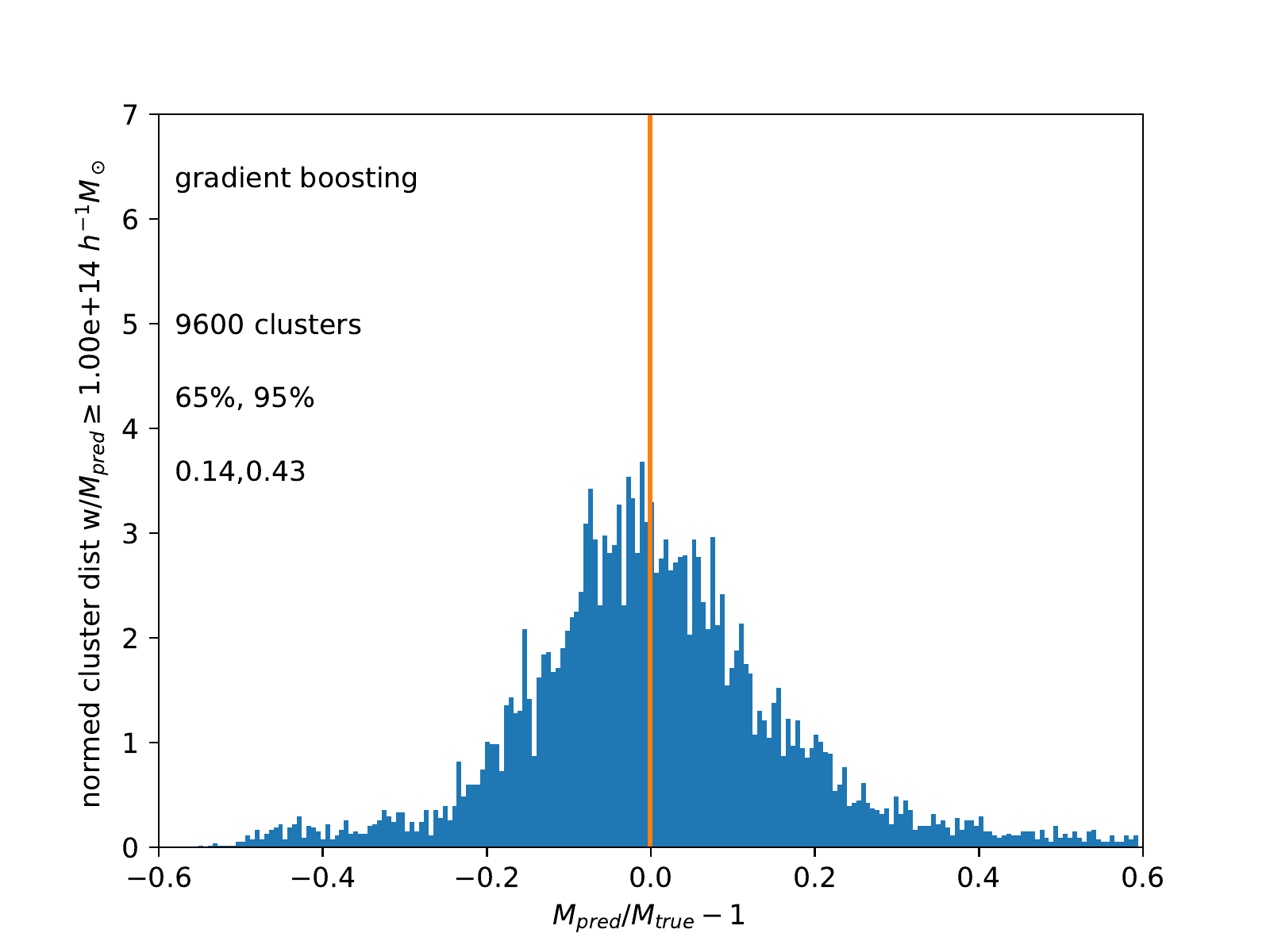}}
\resizebox{3.3in}{!}{\includegraphics{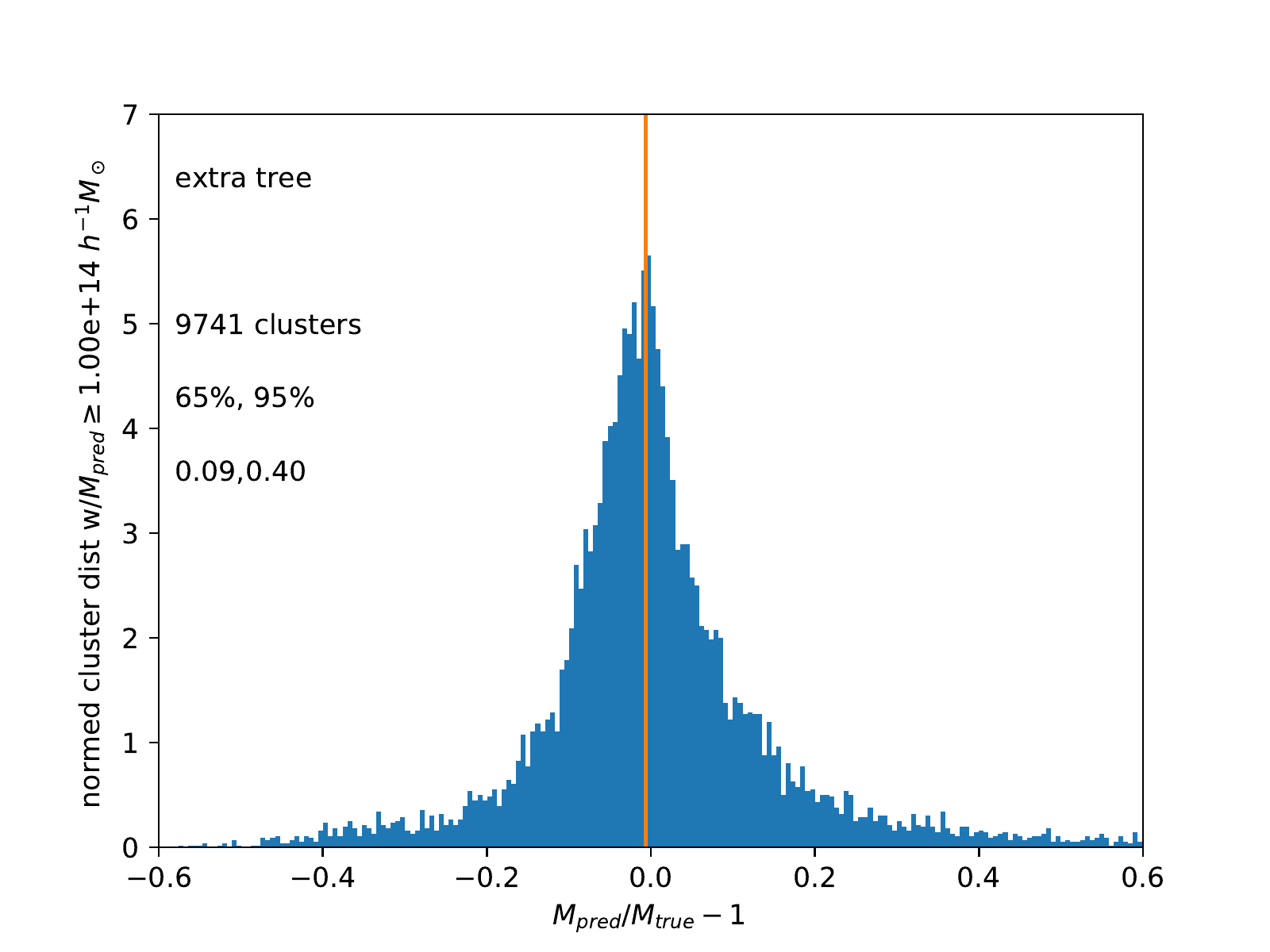}}
\resizebox{3.3in}{!}{\includegraphics{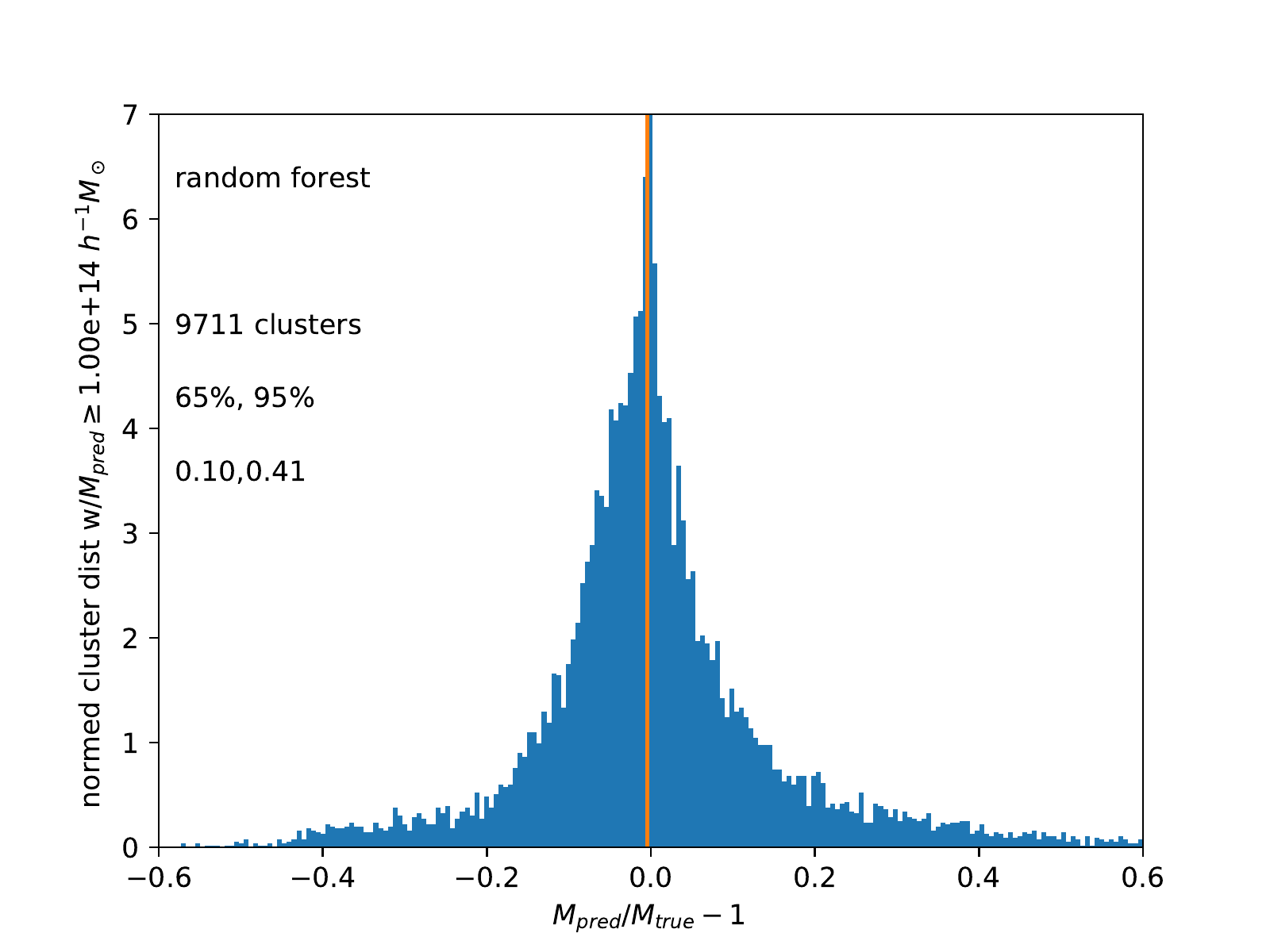}}
\end{center}
\caption{Distribution of $M_{\rm pred}/M_{\rm true} -1$ for clusters
  selected with
  with $M_{\rm pred} \geq 10^{14} h^{-1} M_\odot$, using Gradient Boost, Extra Trees and Random
   Forests, top to bottom.   The 2\% of clusters lying in in the tails of each distribution are omitted.  The values of  $|M_{\rm pred}/M_{\rm true} -1|$ enclosing
  65\% and 95\% of the halos around the median value (vertical line) of $|M_{\rm pred}/M_{\rm true} -1|$ are as listed. The
median $M_{\rm pred}/M_{\rm true}< 1$.  }
\label{fig:errorscatter}
\end{figure}
\begin{figure}
\begin{center}
\resizebox{3.3in}{!}{\includegraphics{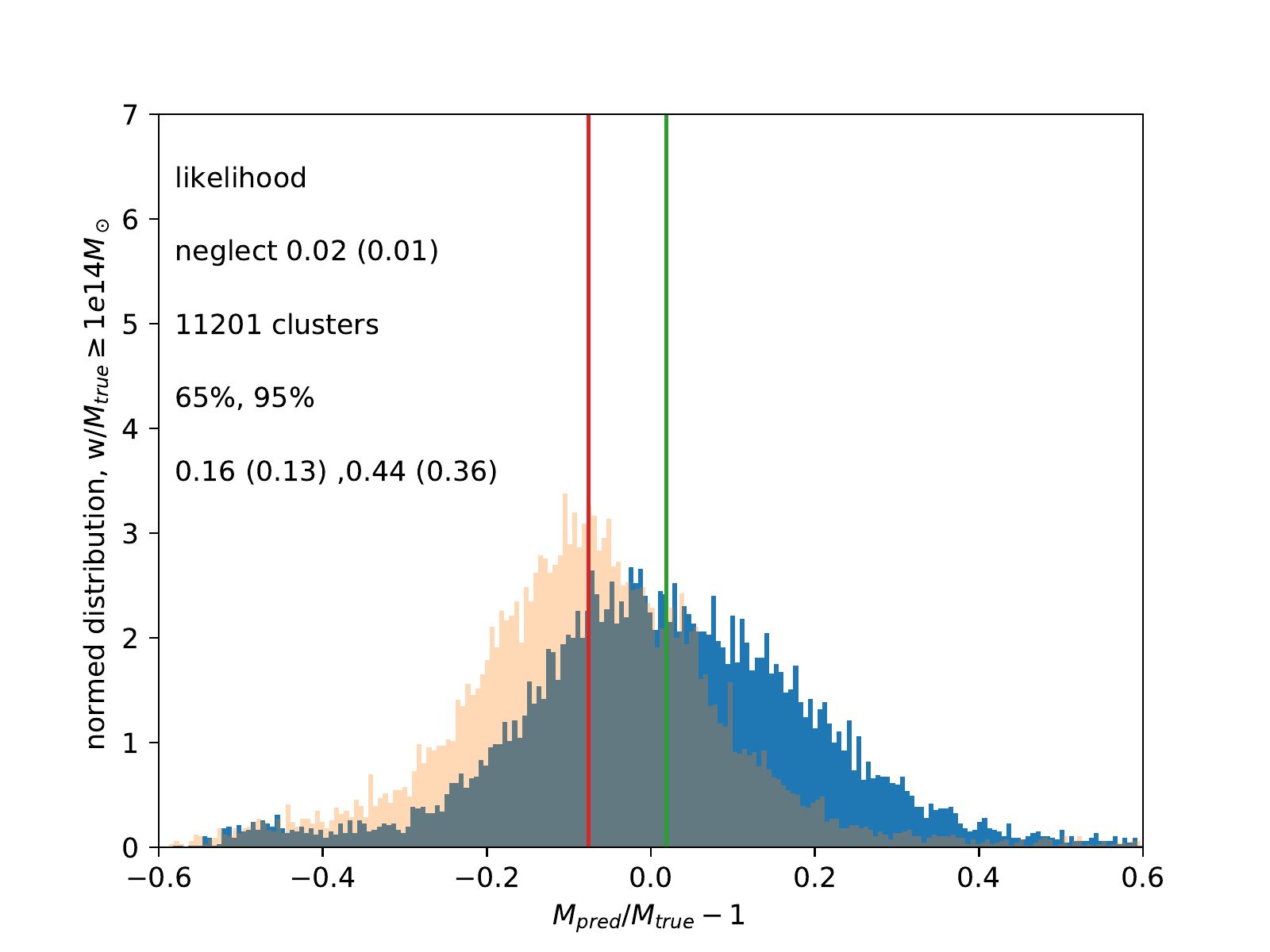}}
\resizebox{3.3in}{!}{\includegraphics{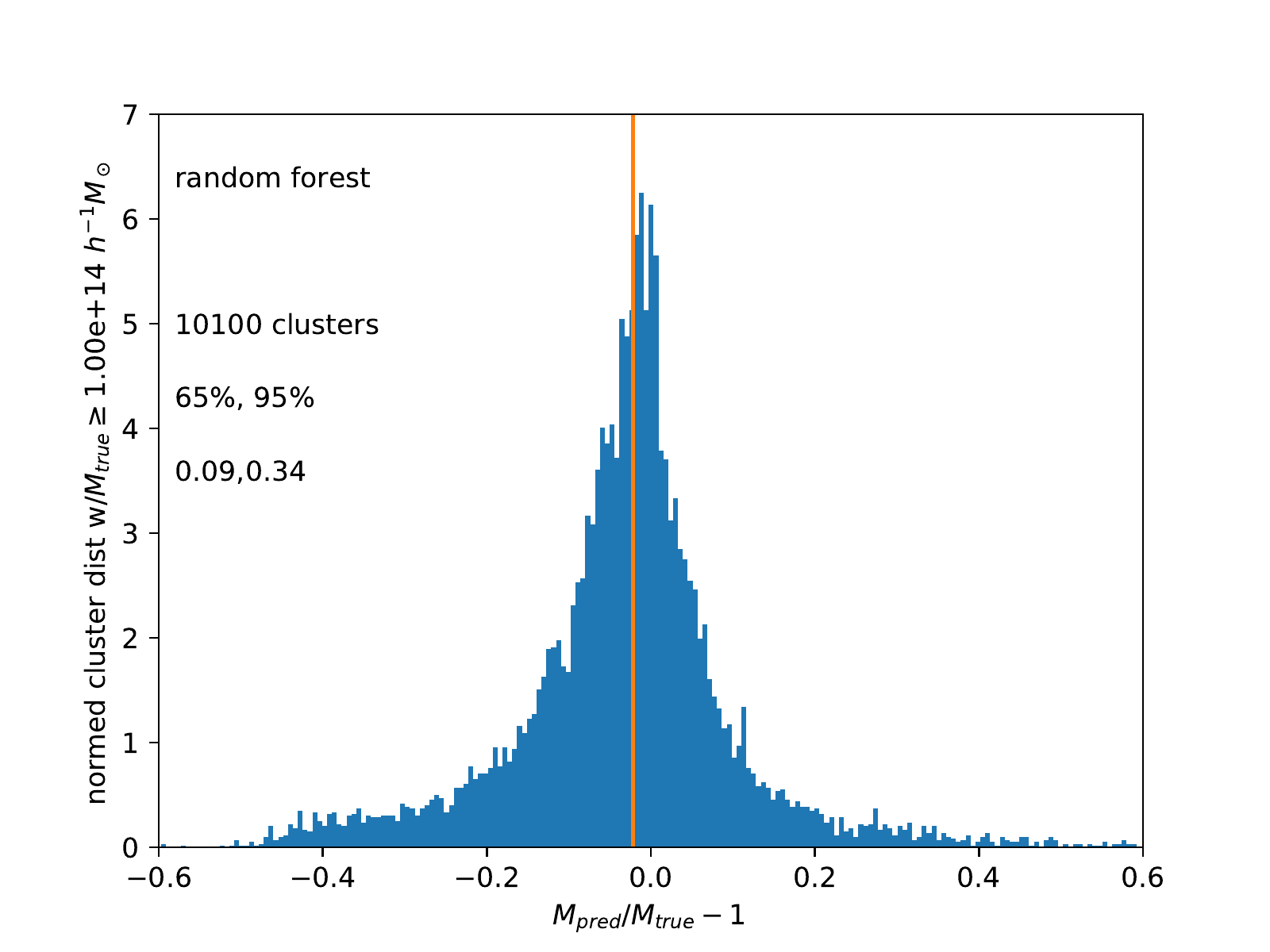}}
\end{center}
\caption{Comparison with a likelihood approach. The cluster sample is selected, for simplicity, on $M_{\rm true} (\geq 10^{14} h^{-1} M_\odot$).  At top are the distributions of $M_{\rm pred}/M_{\rm true} -1$ for clusters using likelihood methods based upon the two scaling relations in Fig.~\ref{fig:observables}. The solid line (dashed line) in Fig.~\ref{fig:observables} corresponds to the darker (lighter) histogram. At bottom is the same quantity for the same sample, using Random Forests to predict mass.
   The 2\% (1\%) (top) and $<$1\% (bottom) of clusters lying in the tails of each distribution are omitted.  The values of  $|M_{\rm pred}/M_{\rm true} -1|$ enclosing
  65\% and 95\% of the halos around the median value (vertical line) of $|M_{\rm pred}/M_{\rm true} -1|$ are as listed. }
\label{fig:errorlikelihood}
\end{figure}

To estimate the scatter in true mass for a given predicted
cluster mass, the distribution in the simulation must be chosen not according to the unobservable true mass, as in Fig.~\ref{fig:corrlns}, but according to
the same selection function as was used to construct the observed sample.
If the simulation can be sampled in the same way as the observations, i.e. with the same cuts on SZ flux $Y_{\rm int}$, etc., and is accurate enough, the true mass distributions can be predicted by combining the observations and the simulation. 
One way to estimate the accuracy of $M_{\rm pred}$ for an observation,  without a particular survey-specific selection function in
hand, is to make a cut in $M_{\rm pred}$, based on Fig.~\ref{fig:corrlns}.  This cut appears to be complete based upon Fig.~\ref{fig:corrlns}, and so to the extent the simulation is accurate and thus to be trusted for the machine learning predictions, it can also be used to estimate the regime of applicability.  For $M_{\rm pred}\geq 10^{14} h^{-1} M_\odot$, it seems all the contributions from the lower mass clusters
$5\times 10^{13} h^{-1}M_\odot$ and above will be captured. Ideally there would be even lower mass halos
  measured with these observations in the simulation, and a larger sample, to confirm
  lower masses are not being
  missed with this cut, but these are not available in the
 current data set.  This $M_{\rm pred}$ cut can
be applied to both the simulation and observations without knowing the
true mass distribution in the observations, and is just a stand-in for a specific but
unknown selection function corresponding to a given survey. 

For  $M_{\rm pred}\geq 10^{14} h^{-1} M_\odot$, $M_{\rm
  pred}/M_{\rm true} - 1$ is shown in Fig.~\ref{fig:errorscatter} for
the three different ML
algorithms also used in Fig.~\ref{fig:corrlns}.  The distributions are not
well fit by
gaussian or lognormal distributions, so values of  $M_{\rm
  pred}/M_{\rm true} - 1$ encompassing 65\% and
95\% of the galaxies
around the median are shown. The vertical line is the median $M_{\rm pred}/M_{\rm true}$, which is slightly less than one. 

Extra Trees and Random Forests have narrower scatter in $M_{\rm pred}/M_{\rm true} -1$ than Gradient Boost, and all three histograms
 have tails to large $M_{\rm pred}/M_{\rm true}-1$, comprising the
 2\% of cluster lines of sight which are not shown.
Again, the advantage of this approach is to calculate the mass from
the measurements in combination, all at once.  In principle, errors and scaling relations for the usual approach would
also have to be calculated from a simulation, using distributions such as
are in Fig.~\ref{fig:observables}, with close attention to using the
relevant selection function and its correlations with the multiwavelength measurements.  Unfortunately, the effect of these errors on the mass function as a whole was difficult to estimate without adding many other assumptions, as the sample was only large because it was replicated 96 times using different lines of sight. 

As pointed out by the referee, it might be expected that the tree
methods (Extra Trees and Random Forest) would work better than the
boosting method, as boosts work better with high bias data, and reduce
bias, and trees work better with high variance data, and reduce
variance.  If one is using scaling relations, the mass error for each observable seems to be more dominated by variance than by bias.  

For comparison with scaling relation approaches, a simple likelihood
estimate can also be made. Scaling relations are found by a least
squares fit $M = A (obs)^B$ to the distributions in
Fig.~\ref{fig:observables}.  The direct least squares fit (dashed
line) is terrible, with the huge scatter for the more numerous low mass clusters causing
a bias for the high mass clusters. Thus, a second fit was also used, where the true masses were binned in 20 log $M_{\rm true}$ bins, and then the median mass in each was taken to correspond to the median observable in each. The least squares fit for the nonzero bins in this case produces the solid lines shown in Fig.~\ref{fig:observables}.
 For each observable, each cluster then has a scaling law mass prediction $\vec{M}_{\rm pred} =(M_1,M_2,M_3,M_4,M_5)=(M_{N_{\rm red}},M_{MF},M_{vel},M_{WL \; shear}, M_{SZ})$.  The scatter covariances, 
$ \langle(M_{\rm true} - M_{i})(M_{\rm true}-M_{j}) \rangle = C_{ij}$ are then calculated and used to give the likelihood prediction for the mass of any cluster:
\begin{equation}
M_{pred} = \frac{\sum_{i,j} (C^{-1})_{i,j} M_{j}}{\sum_{i,j} (C^{-1})_{i,j}}
\end{equation}
The ratios $M_{\rm pred}/M_{\rm true} -1$ are shown in Fig.~\ref{fig:errorlikelihood} for both this estimate, and the Random Forests estimate.  For simplicity, a true mass cut, $M_{\rm true} \geq \times 10^{14} h^{-1} M_\odot$, is taken for the sample, both in the likelihood and ML estimates, to avoid the additional complications introduced by observable based mass cuts. The favorable comparison between the out of the box machine learning estimate relative to a similarly simple likelihood estimate suggest that further investigation of the ML learning methods for directly combining observables would be promising.

Comparing these two kinds of likelihood estimates against the machine learning predictions was also done for the sample used for the WCS, which has a bigger box ($250 Mpc/h ^3$), higher resolution ($2048^3$ particles) and slightly different cosmology ($\Omega_m$ = 0.274, $\Omega_\Lambda$= 0.726, $h$=0.7,$n$=0.95, $\sigma_8$ =0.8). This sample has more clusters above $10^14 h^{-1} M_\odot$, 256 to the 117 in the smaller box, but only mock measurements for clusters with true mass down to $10^{14} h^{-1} M_\odot$. Thus, as mentioned earlier, was not used for the other measurements in this note.   
For this cluster sample, both the likelihood and ML fits were not as biased, as the median values of $M_{\rm pred}/M_{\rm true} -1$ were usually close to zero. However, the likelihood fits for the two scaling relations shown in Fig.~\ref{fig:observables} remained broad compared to the ML fits, with half width around the median for 65\% of the clusters at (0.13,0.14) and for 95\% at (0.38,0.36), relative to (0.09) for 65\% and (0.29) for 95\% for the Random Forests fit.   
This is the main point of this note. In \S\ref{sec:discuss}, several possible refinements and improvements in the machine learning approach are listed.

\subsection{Subsets of multiwavelength measurements}
\label{sec:subsets}
One can also study how well mass is predicted by a given subset of the observations (within the context of the idealizations for the WL, SZ and MF measurements mentioned earlier). 
 In Machine Learning, {\it Importance Permutation} \citep[IP, e.g.][]{Breiman01,Strobl07} is a model-agnostic way to determine the relative importance of each observable, i.e., to ask: "how important is each simulated observable when inferring $M_{\rm pred}$ (in these simulations)?"
The IP procedure uses the difference in the prediction accuracy before and after permuting a given predictor observable to quantify its importance. The IP is generally a reliable except for cases where the observables are highly correlated, which will result in an over-estimation of the highest ranked observables \citep{Strobl08}. However, the lowest ranking IP will not be biased as the least correlated observables will also have the lowest IP values. Given the large correlations between observables, the IP values can be used to determine which observables provide the least information. The lowest ranking observable was vel followed by $N_{\rm red}$ for all three machine learning programs, Gradient Boost, Extra Trees and Random Forest. The IP values for vel and $N_{\rm red}$ ranged from 0.004 to 0.006 and 0.02 to 0.05, respectively, compared to roughly factors of two to ten larger IP values for the other observables. These are the raw IP values, which have better statistical properties than the scaled ({\it z-score}) IP values \citep[e.g.,][]{Strobl08}.  Note that these two measurements are also the two with the most complete inclusion of observational contributions to mass scatter.

\begin{figure}
\begin{center}
\hbox{\hspace{-3mm}\resizebox{3.4in}{!}{\includegraphics{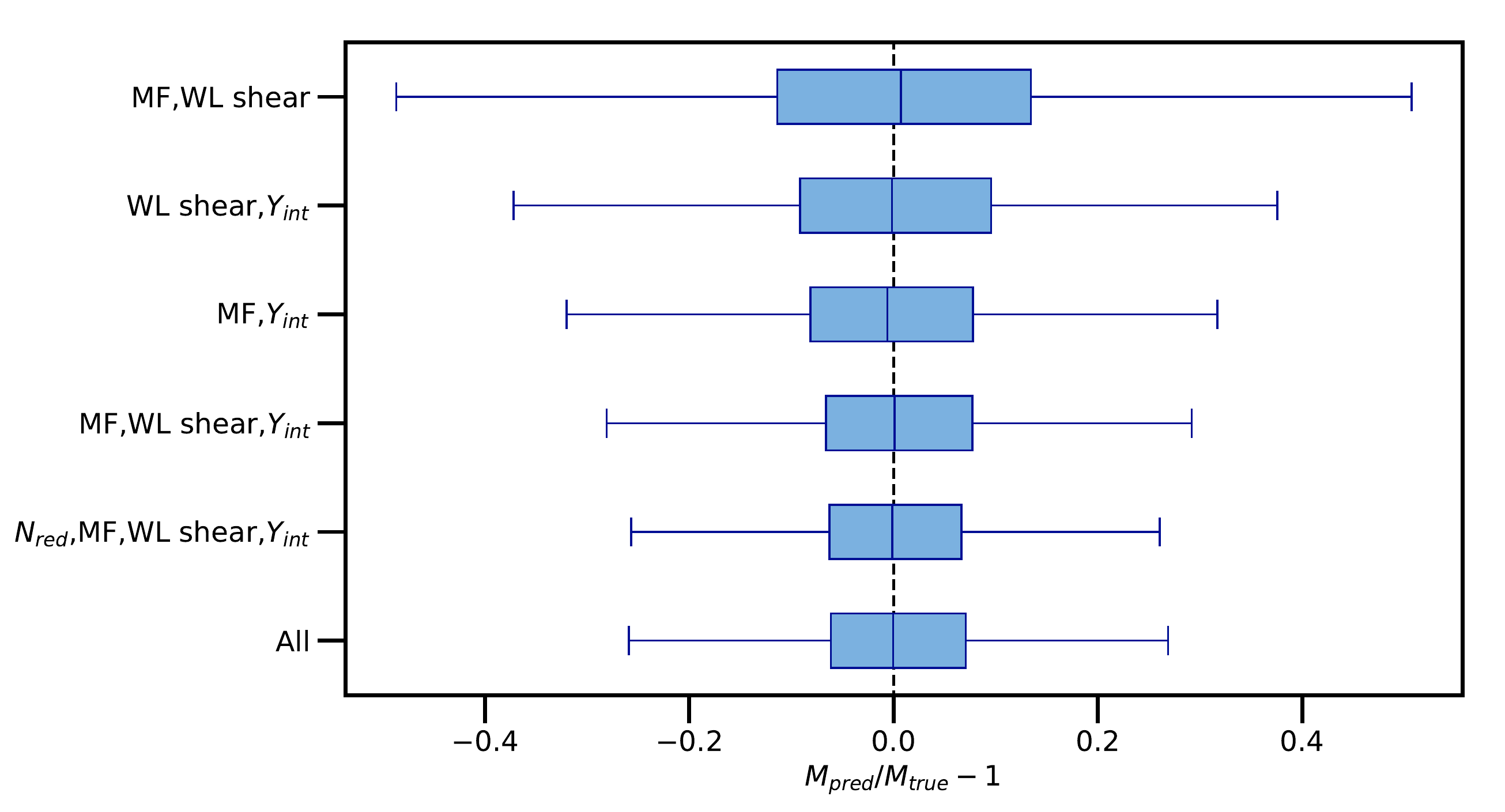}}}
\end{center}
\caption{Mass predictions from subsets of observations.  The median, 25-75 and 0.35-99.65 percentiles of the $M_{\rm pred}/M_{\rm true} -1$ distributions (see Fig.~\ref{fig:errorscatter}) for Extra Trees are illustrated with vertical line inside the box, the edges of the box and the whiskers, respectively. The subsets of observables are listed on the y-axis including the entire set of observables (labeled all) at the bottom for reference. From the bottom, the second and third subsets have the vel and $N_{\rm red}$ observables (with the lowest Important Permutation, or IP, values) cumulatively removed. The upper three subsets show pairs of the remaining three observables: MF, WL shear and $Y_{\rm int}$. The $M_{\rm pred}/M_{\rm true} -1$ distributions upon removing the vel and $N_{\rm red}$ observables show little to no change compared to the all the observables, which is consistent with the IP results that they provide the least information within the assumptions used to create the simulated mock catalogue. Results are similar for Gradient Boost and Random Forest.}
\label{fig:fewerobs}
\end{figure}

Whisker plots in Fig.~\ref{fig:fewerobs} illustrate the impact of removing observables on the $M_{\rm pred}/M_{\rm true}-1$ distributions. Only the Extra Tree distributions are shown in Fig.~\ref{fig:fewerobs}, as results are similar for the other two methods. The baseline is the inclusion of all observables, at bottom. When the lowest IP observables, vel and $N_{\rm red}$, are dropped, the distributions show little to no change, consistent with their low IP values and the statement that they contain the least information for inferring $M_{\rm pred}$. Hence, here the IP values can provide an efficient and effective way to explore the impact of removing observables on the $M_{\rm pred}/M_{\rm true}-1$ distributions. 
The upper three whiskers show the distributions corresponding to pairs of the remaining three observables, MF, WL shear and $Y_{\rm int}$. Of these three, dropping $Y_{\rm int}$ causes the most spread in $M_{\rm pred}/M_{\rm true}-1$.  Combining the whisker plots with the IP results, one finds, for these idealized SZ, WL shear and MF multiwavelength measurements, that $Y_{\rm int}$ is the most important simulated observable for predicting cluster masses.  
 \begin{figure}
\begin{center}
\hbox{\hspace{-6mm}\resizebox{3.8in}{!}{\includegraphics{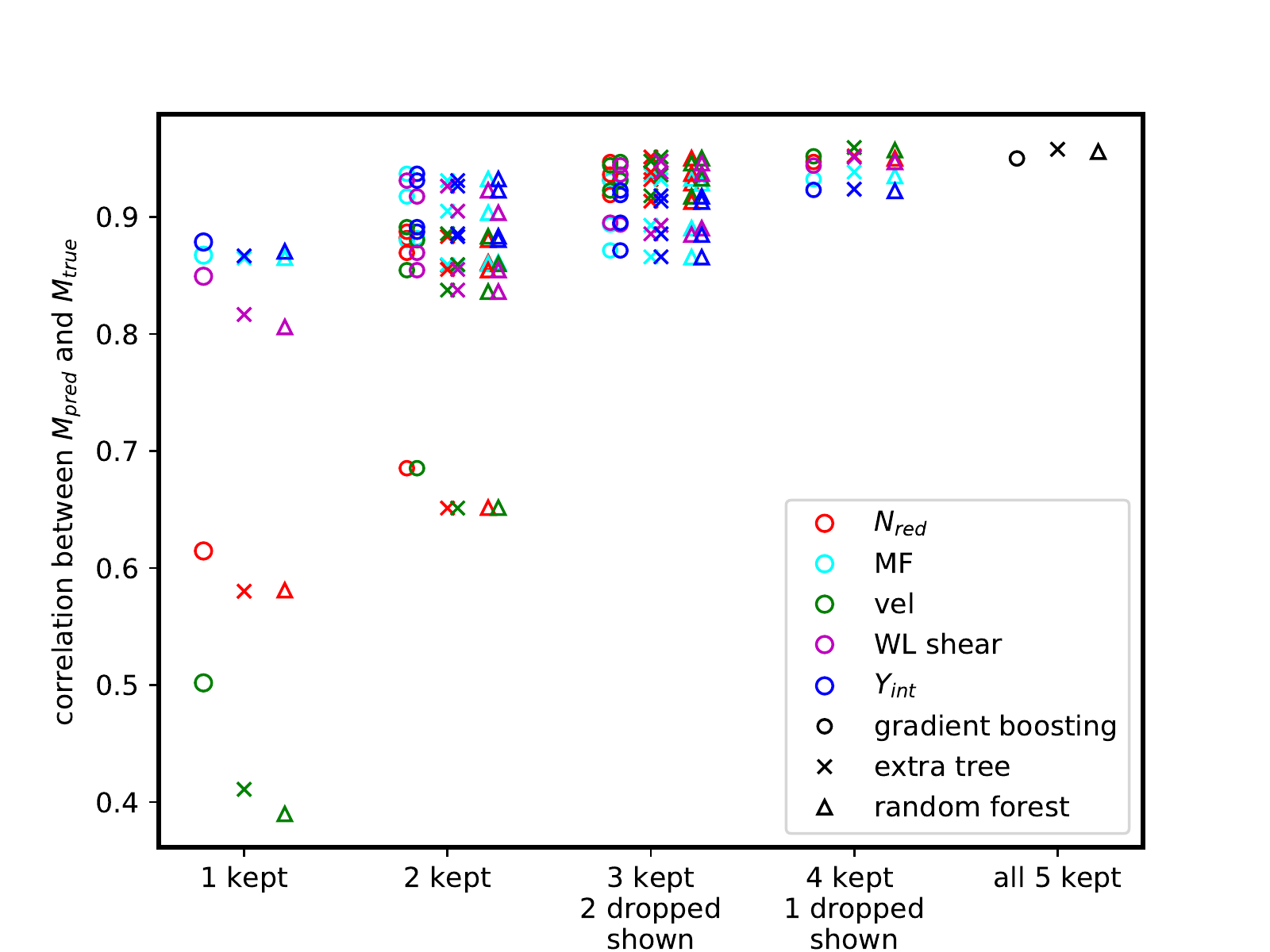}}}
\end{center}
\caption{Correlations between $M_{\rm pred}$ and $M_{\rm true}$ when a subset of the observables are used.  Colors denote which observables are being used (when one or two are used to calculate $M_{\rm pred}$) or neglected (when three or four are used to calculate $M_{\rm pred}$). 
The three shapes correspond to the three machine learning methods as noted, slightly offset horizontally, for readability.  For reference, the $M_{\rm pred}$ to $M_{\rm true}$ correlations when using all measurements, as in Fig.~\ref{fig:corrlns}, are shown in black at far right.}
\label{fig:dropobs}
\end{figure}

A brute force comparison of correlation functions between true and predicted masses gives complementary information.  In Fig.~\ref{fig:dropobs}, 
$M_{\rm pred}$ and $M_{\rm true}$ correlations are shown for all combinations of observables.  The circle, cross and triangle symbols denote the three different machine learning methods, and colors the observation or observations included (for using one or two observables) or the observations or observation dropped (when using three or four observables). These comparisons are based on a cut on true mass.   A second step would be again to impose
the selection function, but this is very survey dependent, and the simulated sample is very small; the idea here is just to give a qualitative sense of how much variation in $M_{\rm pred}$-$M_{\rm true}$ correlations occurs with (combinations of) different methods.

It would be interesting to apply this method to different combinations of proposed observations, for instance to explore survey strategy tradeoffs in the context of other observations.  For such a study to be reliable, of course, the specific survey dependent systematic errors in all of the measurements should be included as much as possible, and the idealizations removed as much as possible.  

\subsection{Simulation validation}
\label{sec:validate}
The accuracy of $M_{\rm pred}$ from machine learning as proposed here depends on the training simulations sufficiently capturing the observables as a function of true mass, the correlations between the observables, and the effects of the selection function.
(As mentioned earlier, other mass estimates using scaling relations and correlations derived from simulations depend upon the accuracy of the
simulations as well.  In the case where the
  correlations are taken out by marginalization, the form of the
  relation is required, including the variables upon which it depends, rather than the actual numbers.)  
  
  Simulations and observations
can be tested directly for agreement of some properties, such as number counts of clusters with a certain 
$Y_{\rm int}$ or richness or other observational quantities.
 The hope is to find tests which, if passed, would suggest that the simulation is also accurate for properties which are not directly measurable, such as correlations used by ML, or mass scatter correlations, used in likelihood estimates.  Both of these involve the not directly observable true mass.
\begin{figure}
\begin{center}
\resizebox{3.3in}{!}{\includegraphics{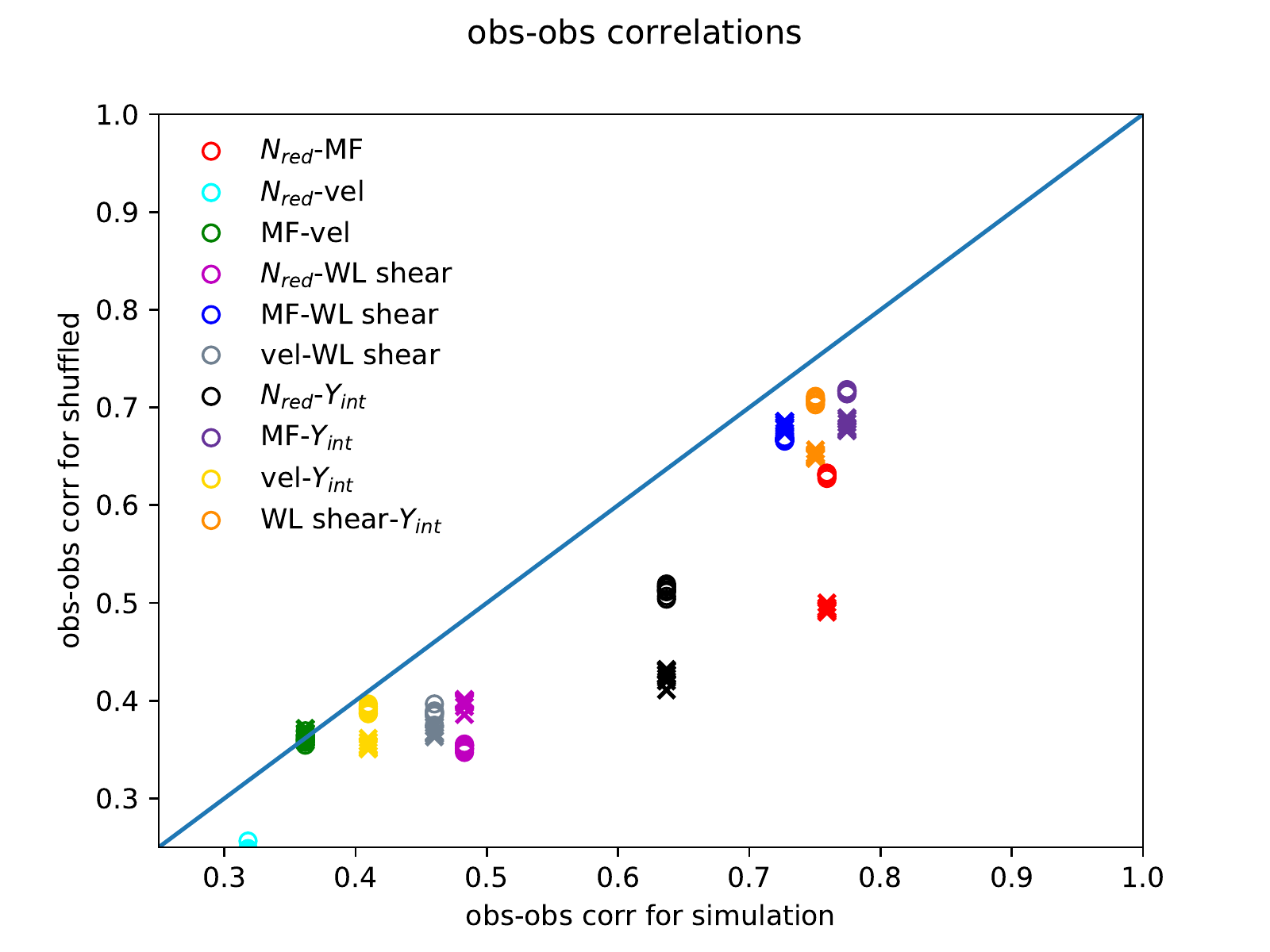}}
\resizebox{3.3in}{!}{\includegraphics{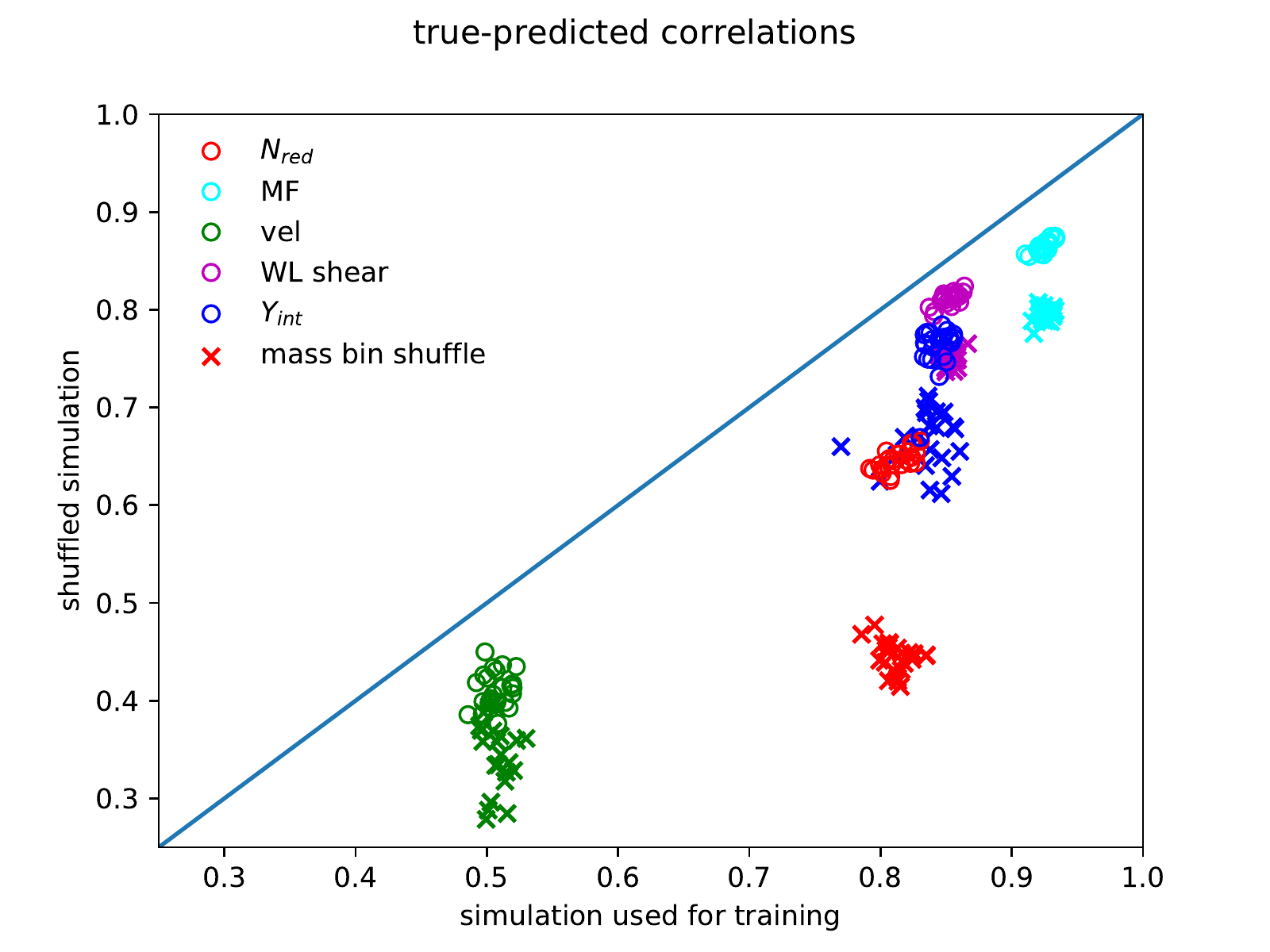}}
\end{center}
\caption{Distinguishing between a data set with the same number counts for observables (as in Fig.~\ref{fig:observables}) but different correlations due to shuffling. Two simulations are considered, the original one, and one where cluster properties are shuffled either between clusters
 within 0.03 dex bins for true mass (crosses) or only between lines of sight for each individual cluster
(open circles).   Top: correlations between pairs of observables; the correlations for observables for the shuffled clusters are, as expected, lower than the unshuffled ones.  Bottom: Correlations between predicted and true values of each of the five
  observables, separately, when predicted by the other four (x-axis), versus correlations between predictions and true values when the machine learning algorithm trained on the original simulation is tested on each of the two shuffled simulations instead.  All 3 machine learning methods are included.  The correlation relative to the original simulation degrades more strongly if cluster properties are shuffled within a mass bin, not only between lines of sight.
    See text for more discussion.  Clusters have observed WL shear $\sigma_{\rm lens}$ $\geq$ 530 km/s in all cases.}
\label{fig:corr2way} 
\end{figure}

\begin{figure}
\begin{center}
\resizebox{3.3in}{!}{\includegraphics{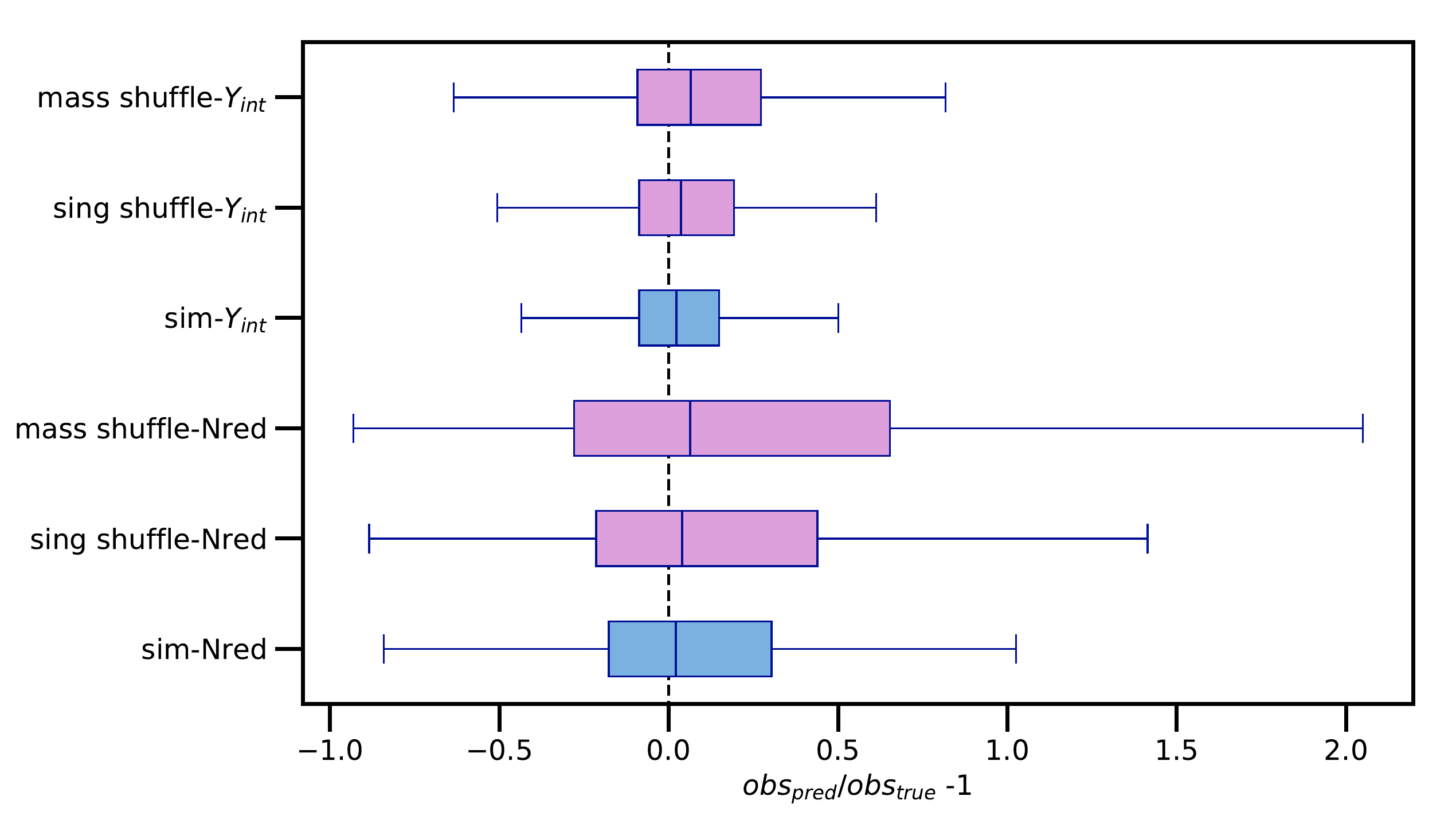}}
\end{center}
\caption{Distribution of Extra Tree pred/true $Y_{\rm int}$ -1 (top 3 whiskers) and pred/true $N_{\rm red}$
  richness -1 (bottom 3 whiskers) for all clusters in sample, predicted in each case
  from the other 4 observables.  Again, as in Fig.~\ref{fig:dropobs}, the median, 25-75 and 0.35-99.65 percentiles are illustrated with vertical central lines, the box edges and the whiskers, respectively. The blue box refers to the training simulation, while the shuffled sightline box (middle of each triplet) and shuffled in 0.03 dex mass bins box(top of each triplet) are in plum.
    Note that ``true'' means the value of
$Y_{\rm int}$ or $N_{\rm red}$ richness measured in the simulation or observation, respectively.  The sample selection function is WL shear $\sigma_{\rm lens} \geq 530 km/s$.}
\label{fig:predhists}
\end{figure}
One test which an accurate simulation must pass is to capture correlations between observed properties, for instance requiring a simulation (with the same selection function) validate the correlations estimated in \citet{Far19}.  This is used regularly to test galaxy formation simulation models, e.g., in \citet{AgaDavBas18}. 

Another angle to try to test simulated correlations between observables and the true mass is to use the machine learning encoding of joint relations between different observables, rather than only the pairwise correlation function of observables above. 
Specifically, one can use machine learning to predict one observable given the others, by training on all observables but one in the simulation (instead of using all observables to predict the observationally inaccessible $M_{\rm true}$ as in \S\ref{sec:masstests}). If the simulation 
is sufficiently accurate, the machine learning relations between observables, derived from mock simulation observations, should also hold for actual observations.  Here, the true and predicted (from ML) simulation observable pairs, for instance, for $Y_{\rm int}$, can be compared to the true vs. predicted observation observable pairs for $Y_{\rm int}$, training and testing on the remaining four observables in the simulation, and then also testing on the observables from an observation. Crucially, predictions for both the simulation observables and the observational counterparts are based on the simulation training set.\footnote{After submission of this paper, we learned of training on one simulation and testing on another as a test being used in simulations to recover star formation histories from spectra \citep{Lov19}.}
If the simulation is close to the observations, the relation of the predictions and true values in the mock observations should be similar to the relation of the predictions and true values in the actual observations. 

As the interest is in the case where the simulation matches the number counts (a first check that a reasonable training simulation will pass) but does not match all the correlations between observables, and 
 given that there is no simulation that agrees with all observations, shuffled line of sight variants of the original simulation are taken as "observation".  These shuffled "observations" have the same number counts as the original simulation, but not the same correlations between observables.  Two shufflings are considered: 
 \begin{itemize}
 \item  shuffling different lines of sight for each observable, separately for each individual cluster, and 
 \item shuffling values of observables between lines of sight and
   between clusters within 0.03 dex bins in halo mass.  
 \end{itemize} The latter shuffling not only mixes lines of sight but other (possibly isotropic) effects, such as local density, which might vary between clusters sharing the same mass.\footnote{We thank M. White for emphasizing this point.} Each kind of shuffling is done 8 separate times, to give 16 samples for comparison with the original.
To approximate a survey selection function, a minimum cut on WL shear $\sigma_{\rm lens}$ of 530 km/s (which seems to reject most low mass halos, see Fig.~\ref{fig:observables}) is made.
Shuffling is done after the WL shear cut, to preserve number counts. The selection function cuts out many more clusters than the $M_{\rm pred}$ selection in \S\ref{sec:masstests}, however, imposing an appropriate $M_{\rm pred}$ cut for the shuffled samples seemed to require more steps and assumptions.

Figures \ref{fig:corr2way} and \ref{fig:predhists} illustrate these two kinds of comparisons between simulations and observations: direct correlations between observables, and correlations between ML predicted and true observables, each determined by the other four observables.

In the top panel of Fig.~\ref{fig:corr2way}, correlations are shown between different simulation observables (x-axis) and their counterparts for the shuffled variants ("observations," y-axis); if the two agreed exactly, the points would lie on the diagonal line.  Of course, the observables should always be correlated to some extent as they are both taken to be proxies for true mass; however, if the scatter in observables from true mass is also correlated,  when the mass-observable scatter is modified by fixing mass and shuffling observables, the correlations will change, as is seen.     

The bottom panel of Fig.~\ref{fig:corr2way} compares correlations between machine learning predicted single observables and true observables in the simulation ($x$-axis), and the predicted observables and true values in their shuffled counterparts ("observations," $y$-axis).  The predicted values of each observable is obtained using the other four via machine learning, trained in both cases on 1/10 of the unshuffled simulation sample.  

In both panels, crosses are mass bin shuffled variants, circles are cluster-by-cluster shuffled variants, and colors indicate which observable or observables are involved.  Some pair correlations (observable-observable, top panel) or internally predicted quantities (true-pred, bottom panel) change strongly when the original simulation properties are shuffled, but not all.
With an actual set of observations in hand, these shuffled simulation variants may be useful in helping to bound the expected response to changing correlations between observables (again, to the extent that the simulations are faithful).

The simulation and shuffled variants are also compared via whisker plots of machine learning predicted/true -1 values of the fifth observable, for two observables
 in Fig.~\ref{fig:predhists}.  Extra Trees is shown for this example, as it gave the highest correlations
between true and predicted for the original simulation.  Again, the mass shuffled sample
is more distinguishable from the original sample than the sample with properties shuffled only cluster by cluster, possibly due to local density variations as mentioned above.  

The above examples show how correlations between observables themselves, correlations between true and machine learning predicted observables and predicted/true ratio distributions
could be used to flag disagreements between simulations and
observations (in practice, replacing the shuffled simulation in this example with actual sky observations, but perhaps using simulation shuffling to help bound the expected effects).

\section{Summary and Discussion}
\label{sec:discuss}
In this note, three ideas were tried out on a simulation:
\begin{enumerate}
\item Using machine learning plus a simulation with mock observations to predict halo masses directly from multiwavelength observations, rather than going through scaling relation mass assignments plus correlated scatter estimates.  This produced fairly tight correlations between $M_{\rm pred}-M_{\rm true}$.
\item Comparing the accuracy of machine learning mass predictions for different combinations of multiwavelength measurements, which for this simulation and mock observations gave clear answers as to which combinations have the most constraining power.  These were compared via IP and brute force.  The specific preferred combinations found here are not directly comparable to specific surveys.  Although the $N_{\rm red}$ and vel measurement errors are faithfully represented in this data set, the other three methods benefit from neglecting survey dependent systematics (SZ and WL), using true mass for some scalings (SZ, WL, MF, often a weak effect) and real space (MF).  These systematics should be included, and the measurement approximations dropped to the appropriate extent, in order to do analyses for specific surveys.
\item Testing correlations in simulations (here used to train machine learning) by comparing relations between observables in the simulated mock observations to those in the observations.  Testing is done by correlating observables within the simulated or observational data sets directly or via machine learning.  In the latter case using one training sample to make predictions for both the simulations and the observations.  For the example here, observations with different correlations were generated from the simulation by shuffling the different multiwavelength measurements between lines of sight, either for each cluster separately, or in mass bins.  Differences were seen between the simulation and its shuffled counterparts for almost all variations of these tests. 
\end{enumerate}

 These initial forays give encouraging results.  There are several ways to go further, following up on this promising exploration.  More extensive simulated data sets could allow all of these ideas to be more thoroughly tested, refined and applied
 and could have:
\begin{itemize}
\item  larger volume and thus more clusters \citep[there are larger simulations in the literature with
  with several different observables, e.g.,][]{Ang12}. 
  
  A larger sample might also allow using
  fewer lines of sight per cluster, to test for and if need be get rid of possible effects introduced by 
 related lines of sight in the small sample used here.  In particular, the small number of distinct high mass halo measurements makes it difficult to estimate the effect of the machine learning mass errors on the mass function.
    
    Also, a larger sample might allow a more refined comparison of correlations, for example, one could consider smaller bins in some observable, for instance, clusters which lie in a specific WL shear bin, more similar to \citet{Far19}, rather than just above a minimum value as done here in \S\ref{sec:validate}.
    
    Lastly, the small number
    of distinct true masses in this sample may also have some effect on ML 
    success.  {\tt DecisionTreeRegressor} was also explored, for instance, but had an unreasonably high success rate, seemingly due to matching clusters directly to the exact masses found in the training set.  The small number of true masses doesn't seem to be an issue in
    the machine learning methods shown here, but a 
    broader sampling of the space of possible clusters would be better.

\item more simulated rather than ``painted on'' physical properties, for instance, using
  hydrodynamics to capture more of the baryonic effects during evolution \citep[e.g.,][]{ShiNagLau16,Hen17}, and more generally, improvements of mock observations.
\item more cluster observables, for instance, other richness
  estimators could be used, such as many of the galaxy based cluster mass
  estimators which have been compared in \citet{Old15}.  The
    velocity information used in \citet{Nta15,Nta16,Nta17,Ho19}
    could certainly be added, or X-ray, which plays an important role in many cluster studies. X-ray luminosity and X-ray temperature measurements are included in the machine learning cluster mass estimates for a 357 cluster sample by \citet{ArmKayBar19}, along with mean, standard deviation, kurtosis and skewness of angular positions, stellar masses, velocities, and angular positions plus velocities of galaxies in a cylinder, number of galaxies, galaxy distribution axis ratio and SZ flux within 5 $R_{200c}$. Some of these would complement well the observables used here.
\item  more simulated observations of smaller mass halos
and other possible configurations which might also pass the cluster selection cuts
(depending upon the cluster finding algorithm,
i.e., sample selection).
\end{itemize}

For mass predictions, improved machine learning algorithms are likely possible, as these were just "out-of-the-box".  Extensive comparisons of algorithms could be reasonably expected to yield even better results.  Ideally, such comparisons would be on data sets that are closely tuned to specific observational surveys (such as using specific cluster selection functions, or best guess estimated errors for, for instance, photometric redshifts or other measurement ingredients).  In addition, it is possible that some preprocessing of the data would give more successful results, such as using observables raised to some power, thus combining some of the physics from the scaling relations with the machine learning inferences from the simulation data.

In addition, it would be good to know if there are better ways than
simply sampling, as is done here, to estimate the error on $P(M_{\rm
  true}|M_{\rm pred})$ as found by these machine learning
approaches (such as using a``prediction interval'').  For the whole sample, the
distribution of predictions from a simulation is a good first estimate of these probabilities,
assuming the simulation is a representative sampling, but it would be very good to have a
way to analyze smaller distributions of observed clusters, down
to, for example, using extensive multiwavelength measurements of a
single cluster to estimate the likelihood that the cluster came from the same distribution
as that of a given simulation.   The scatter in  $P(M_{\rm
  true}|M_{\rm pred})$  in Fig.~\ref{fig:errorscatter} seems fairly small, but in
practice would have to calculated with a
simulation with a better sampled population as described above, and the errors above would need to be combined
with the uncertainties going into the simulation (see, for example,
estimates in \citet{ShiNagLau16} for an enumeration of some of these). 

In testing the simulations for accuracy, perhaps
further
  information (such as ``importance'') could be used to inform the simulation if it does not sufficiently match observations, and more sophisticated definitions of ``match" could be used.  It would also be
  good to incorporate the observational measurement error bars into the simulation tests,
  perhaps by using a distribution of observational input measurement values weighted by their likelihood.  As the errors are different in
  observations and in the simulations, it seems that training on observations might also be possible (to predict one observable from the others), but one of the two training sets is likely preferable.
  In addition, the shuffling test here is similar to the analysis made setting the correlations to zero in \citet{ShiNagLau16}; that is, the two sorts of correlations are the ones in the simulation and their decorrelated counterpart.  It is not clear how a different correlation (rather than a decorrelating shuffling) would appear in these tests.

To summarize, matching each
observable to a mass and then combining mass predictions relies upon
calibrating each mass prediction separately, while combining the
observables first, as suggested here, relies upon a joint calibration from a simulation
capturing all observed properties.  This machine learning based approach requires a more detailed simulation,
but as some of the underlying physics is shared between observables,
this approach may have the benefit of including more of the true correlations.  Another possibility is to go even further with machine learning, to cosmological parameters, skipping the calculation of cluster masses along the way.\footnote{We thank G. Holder for this suggestion.}  Both approaches are incorporating physical assumptions
about clusters and their masses, but in different ways, giving a more complete understanding when considered together. 

\section*{Acknowledgements}
We thank M. White for the simulations and discussions, G. Evrard
for suggestions on an early draft, and the referee and editor for many helpful questions and suggestions.   The simulations  used  in  this  paper  were  performed  at  the  National Energy Research Scientific Computing Center and the Laboratory Research Computing project at Lawrence Berkeley National Laboratory.


\begin{thebibliography}{99.}
\bibitem[{{Adhikari, Dalal \& Chamberlain}(2014)}]{AdhDalCha14}
Adhikari, S., Dalal, N., Chamberlain, R.T., 2014, JCAP, 11, 19
\bibitem[{{Agarwal, Dave \& Bassett}(2018)}]{AgaDavBas18}
Agarwal, S., Dave, R., Bassett, B.A., 2018, MNRAS, 478, 3410

\bibitem[{{Allen, Evrard \& Mantz}(2011)}]{AllEvrMan11}
Allen, S.W., Evrard, A.E., Mantz, A.B., 2011, ARA\&A, 49, 409

\bibitem[{{Angulo et al}(2012)}]{Ang12}
Angulo, R. E., Springel, V., White, S. D. M., Jenkins, A., Baugh,
C. M., Frenk, C. S., 2012, MNRAS, 426, 2046

\bibitem[{{Armitage, Kay \& Barnes}(2019)}]{ArmKayBar19}
Armitage, Thomas J., Kay, Scott T., Barnes, David J., 2019, MNRAS, 484, 1526

\bibitem[{{Breiman}(2001)}]{Bre01}
Breiman,L., 2001, Machine Learning, 45, 5

\bibitem[{{Borgani \& Kravtsov}(2011)}]{BorKra11}
Borgani S., Kravtsov A., 2011, Adv. Sci. Lett., 4, 204

\bibitem[{{Breiman}(2001)}]{Breiman01}
Breiman L., 2001, Machine Learning, 45, 5

\bibitem[{{Cole \& Kaiser}(1988)}]{ColKai88}
Cole, S., Kaiser, N., 1988, MNRAS, 233, 637
\bibitem[{{Davis et al.}(1985)}]{DEFW}
Davis M., Efstathiou G., Frenk C.S., White S.D.M., 1985, ApJ, 292, 371

\bibitem[{{Efstathiou et al}(1988)}]{Efs88}
Efstathiou G., Frenk C. S., White S. D. M., Davis M., 1988, MNRAS, 235, 715

\bibitem[{{Farahi et al}(2019)}]{Far19}
Farahi, A., et al, 2019, arxiv:1903.08042

\bibitem[{{Friedman}(2001)}]{Fri01}
Friedman, J., 2001, The Annals of Statistics, Vol. 29, No. 5

\bibitem[{{Geron}(2017)}]{Ger17}
Geron, A., 2017, {\it Hands-On Machine Learning with Scikit-Learn \& TensorFlow},Chap 7, Boston: O'Reilly

\bibitem[{{Geurts, Ernst and Wehenkel}(2006)}]{GeuErnWeh06}
Geurts P., Ernst D., Wehenkel L., 2006, Machine learning, 63, 3


\bibitem[{{Gifford, Miller \& Kern}(2013)}]{GifMilKer13}
Gifford,D. , Miller, C., Kern, N., 2013, ApJ, 773, 116

\bibitem[{{Gill, Knebe \& Gibson}(2005)}]{GilKneGib05}
Gill, S.P.D., Knebe, A., Gibson, B.K., 2005, MNRAS, 356, 1327


\bibitem[{{Hastie, Tibshirani \& Friedman}(2009)}]{HasTibFri09}
Hastie, T., Tibshirani, R., Friedman, J., 2009, {\it The Elements of Statistical Learning}, 2nd Edition, New York: Springer

\bibitem[{{Henson et al}(2017)}]{Hen17}
Henson, M. A., Barnes, D. J., Kay, S. T., McCarthy, I. G., Schaye, J.,
2017, MNRAS, 465, 3361 
\bibitem[{{Ho et al}(2019)}]{Ho19}
Ho, M., Rau, M.M., Ntampaka, M., Farahi, A., Trac, H., Poczos, B.,
2018, arxiv:1902.05950

\bibitem[{{Kaiser}(1984)}]{Kai84}
Kaiser N., 1984, ApJL, 284, L9

\bibitem[{{Kamdar, Turk \& Brunner}(2016a)}]{KamTurBru16a}
Kamdar, H.M., Turk, M.J., Brunner, R.J., 2016a, MNRAS, 455, 642

\bibitem[{{Kamdar, Turk \& Brunner}(2016b)}]{KamTurBru16b}
Kamdar, H.M., Turk, M.J., Brunner, R.J., 2016b, MNRAS, 457, 1162

\bibitem[{{Koester et al.}(2007)}]{maxBCG}
Koester B.P., et al., 2007, ApJ, 660, 221

\bibitem[{{Lovell et al}(2019)}]{Lov19}
Lovell, C.C., Acquaviva, V., Thomas, P.A., Iyer, K.G., Gawiser, E., Wilkins, S.M., 2019, arxiv:1903.10457

\bibitem[{{Mo \& White}(1996)}]{MoWhi96}
Mo H. J., White S. D. M., 1996, MNRAS, 282, 347

\bibitem[{{More, Diemer \& Kravtsov}(2015)}]{MorDieKra15}
More,S., Diemer, B., Kravtsov, A.V., 2015, ApJ, 810, 36

\bibitem[{{Navarro, Frenk \& White}(1997)}]{NFW}
Navarro J.F., Frenk C.S., White S.D.M., 1997, ApJ, 490,
493

\bibitem[{{Noh \& Cohn}(2012)}]{NohCoh12}
Noh, Y., Cohn, J.D., 2012, MNRAS, 426, 1829 

\bibitem[{{Ntampaka et al}(2015)}]{Nta15}
Ntampaka, M., Trac, H., Sutherland, D. J., Battaglia, N., Póczos, B.,
Schneider, J., 2015, ApJ, 803, 50


\bibitem[{{Ntampaka et al}(2016)}]{Nta16}
Ntampaka, M., Trac, H., Sutherland, D. J., Fromenteau, S., Póczos, B.,
Schneider, J., 2015, ApJ, 831, 135

\bibitem[{{Ntampaka et al}(2017)}]{Nta17}
Ntampaka, M., Trac, H., Cisewski, J., Price, L. C., 2017, ApJ, 835, 106

\bibitem[{{Old et al}(2015)}]{Old15}
Old, L., et al, 2015, MNRAS, 449, 1897

\bibitem[{{Pedregosa et al}(2011)}]{scikit}
Pedregosa, F., Varoquaux, G., Gramfort, A., Michel, V., Thirion, B., Grisel, O., Blondel, M., Prettenhofer, P., Weiss, R., Dubourg, V., Vanderplas, J., Passos, A., Cournapeau, D., Brucher, M., Perrot, M., Duchesnay, E., 2011, Journal of Machine Learning Research, 12, 2825 

\bibitem[{{Planck collaboration}(2016)}]{Planck16}
Planck collaboration: Ade et al, 2016, A \& A, 594A, 13

\bibitem[{{Ramon-Ceja et al}(2019)}]{Rametal19}
Ramos-Ceja, M.E.,Pacaud, F., Reiprich, T.H., Migkas, K., Lovisari, L., Schellenberger, G., 2019, arXiv:1904.10275

\bibitem[{{Rines et al}(2003)}]{Rin03}
Rines, K., Geller, M.J., Kurtz, J.J., Diaferio, A., 2003, AJ, 126, 2152

\bibitem[{{Rykoff et al}(2008)}]{Ryk08}
Rykoff, E.S., et al., 2008, MNRAS Letters, 387, 28

\bibitem[{{Rykoff et al}(2014)}]{redmapper}
Rykoff, E.S., et al, 2014, ApJ, 785, 104

\bibitem[{{Shin et al}(2019)}]{Shi19}
	Shin, T., et al, 2019, MNRAS, 487, 2900

\bibitem[{{Shirasaki, Nagai \& Lau}(2016)}]{ShiNagLau16}
Shirasaki, M., Nagai, D., Lau, E.T., 2016, MNRAS, 460, 3913

\bibitem[{{Skibba \& Sheth}(2009)}]{SkiShe09}
Skibba R.A., Sheth R.K., 2009, MNRAS, 392, 1080

\bibitem[{{Stanek et al.}(2010)}]{Sta10}
Stanek, R., Rasia, E., Evrard, A.E., Pearce, F., Gazzola, L., 2010, MNRAS, 403, 1072

\bibitem[{{Strobl et al.}(2008)}]{Strobl07}
Strobl C., Boulesteix A-L., Zeileis A., Hothorn T., 2007 BMC Bioinformatics, 8, 25

\bibitem[{{Strobl et al.}(2008)}]{Strobl08}
Strobl C., Boulesteix A-L., Kneib T., Augustin T., Zeileis A., 2008, BMC Bioinformatics, 9, 307

\bibitem[{{Sunyaev \& Zel'dovich}(1972)}]{SunZel72}
Sunyaev R.A., Zel'dovich Ya.B., 1972, Comments on Astrophysics and
Space Physics, 4, 173

\bibitem[{{Voit}(2005)}]{Voi05}
Voit G. M., 2005, Reviews of Modern Physics, 77, 207

\bibitem[{{Wechsler \& Tinker}(2018)}]{WecTin18}
Wechsler, R.H., Tinker, J.L., 2018, ARA\&A, 56, 435

\bibitem[{{White}(2002)}]{TreePM}
White M., 2002, ApJS, 143, 241

\bibitem[{{White, Cohn \& Smit}(2010)}]{WCS}
White, M., Cohn, J.D., Smit, R.,
2010, MNRAS, 408, 1818

\bibitem[{{Yang et al}(2007)}]{Yan07}
Yang, X., Mo, H.J., van den Bosch, F.C., Pasquali, A., Li, C., Barden, M., 2007, ApJ, 671, 153
\end{thebibliography}
\end{document}